\documentstyle[11pt,fleqn]{article}
\newcommand{\N}{\mbox{$I\!\!N$}}             
\newcommand{\R}{\mbox{$I\!\!R$}}             
\newcommand{\C}{\mbox{$I\!\!\!\!C$}}         

\begin{document}


\hfill{\sl preprint -  }
\par
\bigskip
\par
\rm


\par
\bigskip
\LARGE
\noindent
{\bf Local $\zeta$-function  techniques vs point-splitting procedure: 
a few rigorous results}
\par
\bigskip
\par
\rm
\normalsize



\large
\noindent {\bf Valter Moretti}\footnote{On leave of absence 
from the European Centre for Theoretical
Studies in Nuclear Physics and Related Areas,\\
 I-38050 Villazzano (TN), Italy. }

\large
\smallskip

\noindent 
Department of Mathematics, Trento University and Istituto Nazionale di 
Fisica Nucleare,\\
Gruppo Collegato di Trento,  I-38050 Povo (TN), Italy.\\
E-mail: moretti@science.unitn.it

\large
\smallskip

\rm\normalsize



\par
\bigskip
\par
\hfill{\sl May 1998}
\par
\medskip
\par\rm



\noindent
{\bf Abstract:}
Some general properties of local $\zeta$-function procedures to  
renormalize some quantities in $D$-dimensional (Euclidean) 
Quantum Field Theory in curved background  are rigorously discussed
for positive scalar operators $-\Delta + V(x)$ in general closed 
$D$-manifolds, and a few comments are given for nonclosed manifolds too.
A general comparison is carried out with respect to the more known 
point-splitting procedure concerning the effective Lagrangian and the
field fluctuations. It is proven that, for $D>1$, the local 
$\zeta$-function and point-splitting approaches lead essentially to the 
same results apart from some differences in the subtraction procedure of 
the Hadamard divergences. It is found that the  $\zeta$ function procedure
picks out a particular term $w_0(x,y)$ in the Hadamard expansion. 
The presence of an untrivial kernel of the operator $-\Delta +V(x)$ may 
produce some differences between the two analyzed approaches.  
Finally, a formal identity  concerning the field fluctuations,
used by physicists, is  discussed and proven within the local 
$\zeta$-function approach. This is done also to reply to recent criticism 
against $\zeta$ function techniques. 

\par

 \rm





\section*{Introduction.}

The $\zeta$ function techniques to regularize the determinant of 
elliptic operators were introduced in Quantum Field Theory 
by  J. S. Dowker and R. Critchley in 1976  \cite{first} and
S.W. Hawking in 1977 \cite{ha}.
Since the appearance of 
these papers, a large use of these techniques has been done by 
physicists, in particular, to compute one-loop partition functions
within  semiclassical approaches to the Quantum Gravity and also 
concerning other related areas \cite{rep,ze,el,ca}.
After the fundamental works cited above, 
many efforts have been spent in studying the black hole entropy
and related problems by these approaches (for recent results 
see \cite{entropy,md,ielmo}).\\
The  {\em local} $\zeta$-function  approach  differs from integrated $\zeta$
function approaches because the former defines quantities which may 
depend on the point on the manifold and thus can be compared with 
analogue regularized and renormalized quantities produced from
different more usual local approaches as the point-splitting  
one \cite{bd,fu}. 
A first step toward these 
approaches was given by R. M. Wald \cite{wa} who considered explicitly
a local $\zeta$ function to regularize and renormalize the local 
effective Lagrangian of a field operator $-\Delta + m^2$.
Similar results, actually in a very formal 
fashion, paying attention to the physical meaning rather than 
the mathematical rigour,  have been obtained
successively for operators $-\Delta + m^{2} +\xi R$ (also considering 
higher
 spin)  employing the so-called "Schwinger-DeWitt expansion".
Within Wald's paper, is conjectured that a method based 
on a local $\zeta$ function should exist also for the stress tensor and
the results  obtained from this method 
should agree with the results obtained via point-splitting 
approaches. A direct rigorous prove of this agreement is contained in the same 
paper concerning the effective Lagrangian in four dimension and
for a motion operator "Laplacian plus squared mass". Formal proofs
concerning the effective Lagrangian in 
more general cases can be found in \cite{bd}. 
Recently, the local approaches have proven to give untrivial results 
if compared with global approaches  based on heat kernel
procedures \cite{articoloZERBINIcognolavanzo,ielmo,md} whenever the 
manifold contains singularities physically relevant. Just as
conjectured by Wald, methods
based on local $\zeta$ function have been found out which are able to
 regularize and
renormalize the one-loop averaged squared field fluctuations \cite{dm} and the 
one-loop stress tensor \cite{moa} directly (anyhow, a first  remarkable 
formal attempt still appeared in \cite{ha}). In all examined cases,  
agreement with the point-splitting technique has arisen as well as, 
sometimes, differences with integrated heat kernel approaches.

This paper is devoted to
 study  the relation between {\em local}
$\zeta$ function approaches and point-splitting techniques in deep and 
within a rigorous mathematical framework.
For this task, in the first section we shall review the basic concepts 
of the heat kernel theory 
 for an operator "Laplacian plus potential" 
 in a closed $D$ dimensional manifold and review some issues related
to the Euclidean path integral.
In the second part  we shall consider the relationship
between heat-kernel, {\em local} $\zeta$ function 
and some physical quantities related to the Euclidean functional integral 
in compact curved manifolds.  We shall study also the relation 
between the point-splitting procedure
and local $\zeta$-function  approach concerning the 
 effective Lagrangian and the field fluctuations.  In particular, we 
 shall prove that the "Green function" generated by the local $\zeta$
 function (defined also when the operator $A^{-1}$ does 
 not exist) has the Hadamard short-distance behaviour
 for any dimension $D>1$. We shall prove that the point-splitting approach 
 gives the same results of  the local $\zeta$ function technique, 
 apart from a different freedom/ambiguity in choosing a particular term 
 in the Hadamard expansion of the Green function of the operator $A$.
 We shall  give also some comments either for the case of the 
 presence of  boundary or a noncompact manifold. 
 In the end, we shall prove that some identities supposed true by
  physicists, concerning the field fluctuations and two-point 
 functions, can be regularized and rigorously proven within
  the local $\zeta$ function approach. 
(We shall find also a simple application of Wodzicki's residue and
Connes' formula.)
  This will be done also to reply to a recent criticism against
 the $\zeta$ function techniques \cite{ev} where it is erroneously 
 argued that similar formal properties do not work within 
 the $\zeta$ function approach.

\section{Preliminaries.}

Within this section, we  summarize some elementary concepts related 
to the  heat-kernel necessary to develop the theory of the  {\em local}
 $\zeta$ function
for differential operators in QFT in curved background.
For sake of brevity, known theorems  or theorems trivially 
generalizable from known result, will be given without explicit proof.
 References and several comments 
concerning these theorems will be anyhow supplied. 
We shall consider almost only compact manifolds.
General references  including topics on $\zeta$ function techniques 
are  \cite{ch} and \cite{sh,de}
which use ``geometrical'' and more ``analytic'' approaches respectively. 
A very general treatise concerning also pseudodifferential operators
is \cite{gi}.

\subsection{General hypotheses.}

Throughout this paper,  ${\cal M}$ is
a Hausdorff, connected, oriented,  $C^{\infty}$ Riemannian $D$-dimensional
manifold. The metric is indicated with $g_{{ab}}$ in local 
  coordinates.  We suppose also that ${\cal M}$ is compact without 
  boundary (namely 
  is ``closed''). We shall consider real elliptic differential operators 
in  the  Schr\"{o}dinger  form "Laplace-Beltrami operator plus 
potential"
\begin{eqnarray}
A' = - \Delta 
 + V \:\: : \:\:
 C^{\infty}(M) \rightarrow L^2({\cal M}, d\mu_g)  \label{d}
\end{eqnarray}
where, locally, $\Delta = \nabla_a\nabla^{a}$,
and  $\nabla$ means the covariant derivative associated to the metric
connection, $d \mu_{g} $ is the Borel measure induced by 
the metric, and $V$ is a {\em real} 
function  belonging to  $C^{\infty}({\cal M})$.
We assume also that $A'$ is bounded below by some $C\geq 0$.\\
All the requirements above on both ${\cal M}$ and $A'$
 are the {\bf general hypotheses} which we shall refer to
 throughout this paper.

 A countable base of the topology  is required
in order to  endow the manifold
with a partition of the unity and make 
$L^2({\cal M},d\mu_g)$ separable. This
allows the use of the integral representation 
Hilbert-Schmidt operators. In our case, the requested topological property
follows from the compactness and the Hausdorff property.

Concerning the operators $A'$ defined above,  we notice that 
they are symmetric on $C^{\infty}({\cal M})$
 and admit self-adjoint extensions since they commute with the 
 anti-unitary complex-conjugation operator in $L^2({\cal M}, d\mu_g)$ 
\cite{rs}. In particular,  one may  consider  the so-called  
Friedrichs self-adjoint extension $A$ of $A'$ \cite{rs} which, as is well 
known, is bounded from below by the same bound of $A'$. 
A sufficient conditions which  assure $A'\geq 0$,
 (see Theorem 4.2.1 in \cite{de}) is the existence of  a strictly positive
 $C^2({\cal M})$ function $\phi$  such that,  everywhere,
\begin{eqnarray}
\phi(x) V(x) - \nabla_{a}\nabla^{a} \phi(x) \geq 0\: \label{phi} 
\end{eqnarray}
In this case $A'$ is bounded below by
$C = \inf_{x\in {\cal M}} \{ V(x) - \nabla_{a}\nabla^{a} \phi(x)/\phi(x)\} 
\geq 0$. Notice that this condition may hold also
for $V$ nonpositive and  the simplest 
sufficient condition, $V(x)\geq 0$ everywhere, is a trivial subcase, as well.

Concerning  {\bf Theorem 1.3} and successive ones,
we shall suppose  also that the 
{\em injectivity radius} $r$ of ${\cal M}$ is strictly 
positive. The {\em injectivity radius} is defined as 
$r = \inf_{p\in {\cal M}} d(p, C_{m}(p))$
where the {\em cut locus} $C_{m}(p)$ is the set of the union
 of the {\em cut point} of $p$ along all of geodesics that start from $p$,
 $d$ is the geodesical distance \cite{ch,mpg}.
The function $(x,y) \mapsto d^{2}(x,y)$ is everywhere
 continuous in ${\cal M}\times{\cal M}$ and, furthermore, the relevant fact 
for the heat kernel expansion theory
 is that, for  $r>0$, whenever $d(p,q) < r $,
 there is a local chart corresponding to a 
normal coordinate system (the  {\em exponential map})
centered in $p$ (resp. $q$) which contains also the point   $q$ (resp. $p$).
Within this neighborhood, the function $x \mapsto d^{2}(p,x)$ 
(resp. $x \mapsto d^{2}(x,q)$)  is also $C^{\infty}$. Moreover, by this result
and  employing Sobolev's Lemma \cite{ru},  one finds 
that
the function $(x,y)\mapsto d^{2}(x,y)$ belongs to $C^{\infty}
(\left\{  (x,y) \in {\cal M}\times {\cal M} \:|\: d(x,y) < r 
\right\})$.

Sufficient conditions for having $r>0$ are found in Chapter 13 of \cite{mpg}. 
In particular, a strictly positive upper bound $K$ for the sectional 
curvature of a compact manifold is sufficient to have $r>0$ since
 $r\geq \pi/\sqrt{K}$.
Notice also that, for instance, a 
Riemannian manifold symmetric under   a Lie group of isometries involves $r>0$ 
trivially.

As a general final remarks, we specify that, throughout this paper,
 "holomorphic" and  "analytic" are synonyms and
 natural units  $c = \hbar =1$ are employed.

\subsection{The physical background.}

All quantities related to $A'$ we shall consider, for $D=4$,
 appear in (Euclidean) QFT in curved 
background and concern the theory of quasifree scalar fields.
In several concrete cases of QFT, the form of $V(x)$ 
is $ m^2 + \xi R(x) + V'(x)$
where $m$ is the mass of the considered field, $R$ is the scalar curvature of 
the  manifold, $\xi$ is a real parameter and $V'$ another smooth function
not dependent on $g_{ab}$.
All the physical quantities we shall consider are formally 
obtained from the Euclidean functional integral
\begin{eqnarray}
Z[A'] :=
 \int {\cal D} \phi \: e^{-\frac{1}{2} 
\int_{\cal M} \phi A' \phi \: d\mu_g}\:.
\label{integral}
\end{eqnarray} 

The integral above can be considered as a partition function of a field
in a particular  quantum state corresponding to a canonical ensemble. 
Often, the limit case of 
vanishing temperature is also considered 
and in that case  the manifold cannot  be compact. 
The direct physical interpretation 
as a partition function should work
provided the manifold has a  Lorentzian section obtained
by analytically continuing some global temporal coordinate $x^{0}=\tau$ 
into imaginary values $\tau \rightarrow it$
and considering (assuming that they exist) 
the induced  continuations of the metric and relevant quantities.
It is required also that $\partial_\tau$ is a global 
Killing field of the Riemannian manifold generated by  an isometry
group $S_{1}$, which 
 can be  continued into a
(generally local) time-like Killing field $\partial_{t}$ in the Lorentzian
section (see \cite{ha} and \cite{wa}). Then one assumes that 
$k_{B}\beta$ is the inverse of the 
temperature of the canonical ensemble quantum state, $\beta$ being the 
period of the coordinate $\tau$. 
Similar interpretations hold for the (analytic continuations of)
 quantities we shall introduce shortly. 
The (thermal) quantum state  which all the theory is referred to is
determined by the Feynman propagator obtained by analytical continuation
of the Green function of the operator $A'$. For this reason the analysis
of the uniqueness of the Green functions of the operator $A'$ is  important.   
A general discussion on these topics, also 
 concerning grand canonical ensemble states can be found in \cite{ha}.

Physicists, rather than trying to interpret
 the integral in (\ref{integral})
as a Wiener measure, generalize the trivial finite-dimensional Gaussian 
integral and re-write the definition of $Z[A']$ as \cite{ha}
\begin{eqnarray}
Z[A'] := \left[\mbox{det}\left(\frac{A'}{\mu
^{2}} \right)\right]^{-1/2}
\label{integral2}
\end{eqnarray} 
provided a useful definition of the determinant of the operator
$A'$ is given.
The mass scale  $\mu$ 
present in the determinant is necessary for dimensional reasons    
\cite{ha}. 
Such a scale introduces an ambiguity which remains in the finite 
renormalization parts of the renormalized quantities and, dealing with
the renormalization of the stress tensor within the
semiclassical approach to the quantum gravity,
it determines the presence of
quadratic-curvature terms in effective Einstein's equations \cite{moa}.
Similar results  are discussed in \cite{waldlibro,bd,fu}
employing other renormalization procedures (point-splitting). 

The theory which we shall summarize in the following has been 
essentially developed 
to give a useful  interpretation of $\mbox{det} A'$, anyhow it has 
been successively developed to study 
several different, formally quadratic in the field,
quantities related  to the functional determinant above.
Some of these are\footnote{In defining the effective action and so on, 
we are employing the opposite sign conventions 
with respect to \cite{moa}, our conventions are the same used in
\cite{wa}.}, where the various symbols "$=$" have to be 
opportunely interpreted,
 the {\em effective action}
\begin{eqnarray}
 S[A']
 = - \ln Z[A'] = \int_{{\cal M}} {\cal L}(x|A')
d\mu_g(x) \:, \label{effective}
\end{eqnarray}
where the integrand is the {\em effective Lagrangian}; the 
{\em field fluctuations}
\begin{eqnarray}
\langle \phi^2(x|A')\rangle = 
 Z[A']^{-1}\int {\cal D} \phi \: e^{-\frac{1}{2} \: 
\int_{\cal M} \phi A' \phi \: d\mu_g}
 \: \phi(x) \phi(x) 
 = \frac{\delta}{\delta J(x)}|_{J\equiv 0}
S[A' + 2J]   \:, \label{formal}
\end{eqnarray}
and the {\em one-loop averaged stress tensor}
\begin{eqnarray}
\langle T_{ab}(x | A') \rangle =
Z[A']^{-1}\int {\cal D} \phi \: 
e^{
-\frac{1}{2} \: \int_{\cal M} \phi A' \phi \: d\mu_g} 
\: T_{ab}(x) 
 = \frac{2}{\sqrt{g}} \frac{\delta}{\delta g_{ab}(x)}
S[A']\:, \label{stress}
\end{eqnarray}
where 
 $T_{ab}(x)$ is the classical stress-tensor 
(in a paper in preparation we analyze  the stress tensor renormalization.)
Recently, some other nonquadratic quantities have been considered in the 
heat-kernel or $\zeta$-function approaches \cite{hu}.

All  quantities in  left hand sides  of (\ref{effective}), 
(\ref{formal}), (\ref{stress}) 
and the corresponding ones in the Lorentzian section 
are affected by divergences whenever one tries to
compute them by trivial procedures \cite{wa,bd,fu}.
For instance, interpreting the functional integral of $\phi(x) \phi(y)$
in (\ref{formal}) as  the Green function of $A'$ (the 
analytic continuation of the Feynman propagator), $G(x,y)$ 
\begin{eqnarray}
\langle  \phi(x) \phi(y) \rangle = Z[A']^{-1}\int {\cal D} \phi  
\:e^{-\frac{1}{2} \: \int_{\cal M} \phi A' \phi \: d\mu_g}\:
 \phi(x)\phi(y)   = G(x,y)\:,
\end{eqnarray} 
the limit  $y\rightarrow x$,
necessary to get $\langle \phi^{2}(x) \rangle$, diverges as is well known.
One is therefore forced to remove {\em by hand} these divergences, this 
nothing but the main idea of the {\em point-splitting procedure}. 
It is worth remarking that  the  {\em definitions} given in terms of
$\zeta$ function  and heat kernel \cite{ha,wa,bd,moa,dm} of the formal
 quantity 
in left hand sides of (\ref{effective}), 
(\ref{formal}), (\ref{stress}) 
contain an implicit {\em infinite renormalization} procedure 
in the sense that these are finally free from divergences.

\subsection{Heat kernel.}

The key to proceed with the $\zeta$-function theory in order to
provide a useful definition of determinant of the operator (\ref{integral2}) 
$A'$ is based upon the following remark.
In the case $A$ is a $n \times n$ Hermitian  matrix with eigenvalues 
$\lambda_{1},\lambda_{2}, \ldots \lambda_{n}$, then (the prime indicates the 
$s$-derivative)
\begin{eqnarray}
\mbox{det} A = \prod_{j=1}^{n} \lambda_{j} = e^{-\zeta'(0|A)}  \label{start} 
\end{eqnarray}
where, we have defined the $\zeta$-function of $A$ as
\begin{eqnarray}
\zeta(s|A) = \sum_{j=1}^{n} \lambda_{j}^{-s}\:.
\end{eqnarray}
The proof of (\ref{start}) is direct.
 Therefore, the idea is to generalize
(\ref{start}) to {\em operators} changing  the sum into a series 
(the spectrum of $A'$ is discrete as we shall see).   
Unfortunately this series diverges at $s=0$ as we shall see shortly.
 Anyhow, it is possible to {\em continue 
analytically} $\zeta(s|A')$ in a neighborhood of $s=0$ and {\em 
define} the determinant of $A'$ in terms of the continued function 
$\zeta(s|A')$. 
 This generalization  requires 
certain well-known mathematical tools and untrivial 
results we go to  summarize. 

First of all, let us give the definition of {\em heat Kernel} for 
operators $A'$ defined above
\cite{ch,de,gi,wa,ha,fu}. Three relevant theorems follows the definition.
\\

\noindent {\bf Definition 1.1.}
{\em Within our general hypotheses on ${\cal M}$ and $A'$,
let us consider, if it exists, a  class of operators
\begin{eqnarray}
(K_t \psi) (x) := \int_{{\cal M}} K(t,x,y|A') \psi (y)\: d\mu_g(y) 
\:\:\:\: t\in (0,+\infty)
\end{eqnarray}
where the integral kernel is required to be 
 $C^0((0,+\infty)\times {\cal M}
\times {\cal M}))$, $C^1((0,+\infty))$ in the variable $t$,
$C^2({\cal M})$ in the variable $x$
and  satisfy
 the ``heat equation'' 
 with  an initial value condition:
\begin{eqnarray}
\left[\frac{d}{dt} + A'_x \right]  K(t,x,y|A) = 0 \label{hk}
\end{eqnarray}
and
\begin{eqnarray}
K(t,x,y|A') \rightarrow \delta(x,y) \mbox{   as   } t\rightarrow 0^+
\label{delta}\:,
\end{eqnarray}
 the limit is understood in a ``distributional'' sense, i.e., once the kernel 
is $y$-integrated on a test function $\psi=\psi(y)$ belonging  
 to $C^0({\cal M})$ and} (\ref{delta}) {\em means
 \begin{eqnarray} 
 \lim_{t\rightarrow 0^{+}}
 \int_{{\cal M}} K(t,x,y|A') \psi (y)\: d\mu_g(y) = \psi(x) 
\:\:\:\: \mbox{ for each  } x\in {\cal M}  \label{-1}
 \end{eqnarray}
   The kernel $K(t,x,y|A')$, if exists, is called 
 {\bf heat kernel} of $A'$}.\\

\noindent {\bf Theorem 1.1.} 
{\em In our general hypotheses on ${\cal M}$ and $A'$}

(a) {\em the set of the operators $K_t$ above defined exists is unique
and consists of self-adjoint, bounded, compact,  Hilbert-Schmidt,
 trace-class operators  
represented by a
$C^{\infty}((0,+\infty)\times {\cal M}\times {\cal M})$
 integral kernel. This is also real, symmetric in $x$ and $y$ }
 {\em and {\em positive} provided  
either $V$ is positive or} (\ref{phi}) {\em is satisfied for some $\phi$ and } 
{\em {\em strictly positive}
  in the case $V \equiv m^{2} \mbox{(constant)} \geq 0$.}

(b) {\em  Moreover
\begin{eqnarray}
K(t,x,y|A') = \sum_{j=0}^{\infty} e^{-t\lambda_j} \phi_{j}(x)
\phi_j^{*}(y) \label{sum 0}
\end{eqnarray}
where the series converges on
 $ [\alpha  +\infty) \times {\cal M}\times {\cal M}$
 absolutely  in uniform sense (i.e. the series of the absolute values
converges uniformly) for any fixed $\alpha \in 
(0,+\infty)$,  
\begin{eqnarray}
0 \leq \lambda_0 \leq \lambda_1 \leq \lambda_2 \:\:... \rightarrow +\infty
\end{eqnarray}
are eigenvalues of $A'$  and 
$\phi_j \in C^{\infty}({\cal M})$ are
 the corresponding normalized eigenvectors, the dimension
of each eigenspace being finite.}

(c)  {\em The class $\{\phi_j| j=0, 1,2, ...\}$ 
defines also
a Hilbertian base of $L^2({\cal M}, d\mu_g)$
and
\begin{eqnarray}
\int_{{\cal M}} d\mu_{g}(x)\:
K(t,x,x|A) = \sum_{j=0}^{+\infty} e^{-\lambda_{j}t} =\mbox{Tr} 
K_{t}  \label{trace}
\:. 
\end{eqnarray} }
All these results are straightforward generalizations of theorems 
contained in  Section 1 of Chapter VI of \cite{ch},  
the convergence properties follow from Mercer's theorem \cite{rn}; 
 see also \cite{de} concerning the issue of the positivity of
the heat kernel and Sections 3 and 4  of Chapter VI of \cite{ch}
concerning the existence of the heat
 kernel under the further hypotheses $r>0$\footnote{The reader has to 
handle with great care the content of \cite{ch}
since, unfortunately, some missprints appear in several statements.
 For instance, (45) in Section 4 Chapter VI
is incorrect due to the presence of the operator $L_x$, this is the reason 
for the introduction of the parameter $\eta$ in our {\bf Theorem 1.3}. 
Moreover, 
{\bf Lemma 2} in \cite{ch} requires 
$F\in C^1$ rather than $F\in C^0$ as erroneously written there.}.
 The existence of the heat kernel
 can be proven without this hypothesis by studying the integral kernel of the 
 exponential of the Friedrichs self-adjoint extension of $A'$ as 
 done in Chapter 5 of \cite{de}.
 By  Nelson's theorem \cite{rs}, using
the class of all linear combinations  of vectors $\phi_{j}$ 
as a dense set of analytic vectors, one proves
 that $A'$ is essentially self-adjoint in 
$C^{\infty}({\cal M})$; this leads to\\

\noindent {\bf Theorem 1.2.}
{\em In our general hypotheses on {\cal M} and  $A'$
defined on the domain $C^{\infty}({\cal M})$,}

 (a) {\em there is only
one self-adjoint extension of $A'$, namely, its closure  $\bar{A'}$, which 
  also coincides with the
Friedrichs self-adjoint  extension of $A'$, $A$;}

(b) {\em  this extension
 is bounded below by the same bound of $A'$ and} 
\begin{eqnarray}   
\sigma(A)=\sigma_{\mbox{\scriptsize p.p.}}(A) =
 \sigma_{\mbox{\scriptsize disc.}}(A) = \{ \lambda_n  | n=0,1,2 
 ...\}\nonumber \:;
\end{eqnarray}

(c)  {\em in the usual  spectral-theory sense, for $t\in (0, +\infty)$}
\begin{eqnarray}
K_t = e^{-t A} \label{exponantial}
\end{eqnarray}
{\em  and $\{ K_t | t\in (0,+\infty)\}$ is a strongly continuous 
one-parameter
 semigroup of bounded operators. In particular
  $K_{t} \rightarrow I$ as $t\rightarrow 0^{+}$ 
  in the strong operator topology  
  and  $K_{t}\rightarrow P_{0}$ as $t\rightarrow +\infty$ 
  ($P_{0}$ being the projector onto  $Ker$ $A$)
 in the strong operator topology.  The  limit above holds  also in the sense
   of the uniform 
 punctual convergence whenever $K_{t}$ acts on
  $\psi \in C^{0}({\cal M})$.}\\

From now on, since $A'$ determines $A$ uniquely and 
 $A' = A|_{C^{\infty}({\cal M})}$, we shall omit the prime on  $A'$
almost everywhere.

Generalizing the content of Section 3 and 4  of Chapter VI of \cite{ch}
\footnote{See the previous footnote.} one also 
gets a well-known ``asymptotic expansion'' of the heat kernel for
$t\rightarrow 0^+$ \cite{ch,fu,ca}.\\
 
 \noindent {\bf Theorem 1.3.}
 {\em  In our general  hypotheses on ${\cal M}$  and $A'$,
 supposing also that $r>0$,}

 (a) {\em  for any fixed integer  $ N >  D/2 +2 $ and any fixed real 
$\eta\in (0,1)$
 it holds 
\begin{eqnarray}
K(t,x,y|A) =
\frac{e^{-\sigma(x,y)/2t}}{(4\pi t)^{D/2}}
 \chi (\sigma(x,y)) \sum_{j=0}^N a_j(x,y|A) t^j + 
\frac{e^{-\eta\sigma(x,y)/2t}}{(4\pi t)^{D/2}} t^N O_\eta(t;x,y)  
\label{expansion1}\:,
\end{eqnarray}
where}\\
(1) {\em 
 $2 \sigma(x,y) = d^{2}(x,y)$, $\chi =\chi(u)$ is a  non-negative function in  
$C^{\infty}([0,+\infty))$
 which takes
the constant value $1$ 
for $|u| < r^2/16$ and 
vanishes for $|u| \geq r^2/4$.}\\
(2) {\em $O_\eta$ is a function in
 $C^{0}([0,+\infty)\times{\cal M}\times {\cal M})$ at least, 
and it is such that  the function  
\begin{eqnarray}
 (t,x,y) \mapsto
  \frac{e^{-\eta\sigma(x,y)/2t}}{(4\pi t)^{D/2}} O_\eta(t,x,y) 
\end{eqnarray}
belongs to 
$ C^{\infty}((0,+\infty)
\times{\cal M}\times {\cal M})$.
Moreover, for any positive constant $U_{\eta}$ and for $ 0 \leq t < U_{\eta}$
 $|O_\eta(t,x,y)| < B_\eta t $ holds true 
for a corresponding positive constant $B_\eta$, 
not depending on $x$ and $y$ in ${\cal M}$.}\\
(3) {\em The coefficients $a_{j}(x,y|A)$ 
are defined when $x$ and $y$ belong to $\{ (x,y) \in {\cal M}\times 
{\cal M} \:| \: d(x,y) <r \}$
 and are $C^{\infty}$
therein, in particular $\chi(\sigma) a_{j} \in C^{\infty}({\cal M}\times
{\cal M})$ $(j = 0,1,2 ...)$.}

(b) {\em The $C^{\infty}((0,+\infty) \times{\cal M}\times {\cal M})$ 
functions 
called {\em parametrices} 
\begin{eqnarray}
F_N(t,x,y) = 
\frac{e^{-\sigma(x,y)/2t}}{(4\pi t)^{D/2}}
\chi(\sigma(x,y)) \sum_{j=0}^N a_j(x,y|A) t^j \:\:\:\: N=0,1,2,\:\:...
\end{eqnarray}
for each $N= 0, 1, 2, ... $ fixed,
 satisfy,  working as integral kernel  on functions in $C^{0}({\cal M})$,}
\begin{eqnarray}
F_N(t,x,y) \rightarrow \delta(x,y) \mbox{   as   } t\rightarrow 0^+
\label{delta2} \:.
\end{eqnarray}

 Notice that, the values $r^{2}/4$ and $r^{2}/16$ in the definition of $\chi$
 may be changed, their task is just to make everywhere $C^{\infty}$
 the right hand side of (\ref{expansion1}) also when $x$ is too far from  $y$.
 Concerning the precise form of the  coefficients $a_{j}$,
we have that all $a_j(x,y|A)$
can be obtained \cite{ca} by
 canceling  both $\chi$ and $O_\eta$ out
 and  {\em formally} substituting
the expansion (\ref{expansion1}) with $N =+\infty $ (this limit 
usually does {\em not} exist)
into the heat kernel equation (\ref{hk}) and requiring that the 
coefficients of each $t^j$ vanish separately.  This produces
the set of recurrent differential equations in each normal 
convex neighborhood ${\cal N}_{y}$ 
centered in $y$ and referred to spherical coordinates 
$(x^{a})_{a=1,\ldots, D} =
(\rho,\Omega) \equiv x$ ($\rho$ is the geodesical distance of $x$ from $y$)
 \cite{ch,ca}
 \begin{eqnarray}
 \rho\partial_{\rho}\left(a_{0}(x,y|A) \Delta_{VVM}^{-1/2}(x,y)\right) &=&
 0 \:,
 \label{eq1}\\ 
- \Delta_{VVM}^{-1/2}(x,y)\: A'_{x}
a_{j}(x,y|A)  &=&   
  \rho\partial_{\rho}\left(a_{j+1}(x,y|A)\: \Delta_{VVM}^{-1/2}(x,y)\right)
\nonumber \\
&+& (j+1) \:a_{j+1}(x,y|A)\: \Delta_{VVM}^{-1/2}(x,y) \:\:\: (j>0)
\:,   \label{eq2}  
 \end{eqnarray}
 $\Delta_{VVM}(x,y)$ is defined below. 
These equations
determine
uniquely the coefficients $a_j(x,y|A)$ once one fixes 
$a_0(x,y|A)$ 
 and requires that  $a_j(x,y|A)$ ($j=0,1,2 \ldots$) is bounded 
as $x\rightarrow y$. \\
To assure the validity of (\ref{delta2}), it is sufficient to
requires that $a_0(x,y|A)$ which satisfies (\ref{eq1}) is smooth and  
$a_0(x,y|A) \rightarrow 1$ as $x\rightarrow y$ not depending  on $\Omega$. 
Then (\ref{eq1})
and  
$\Delta_{VVM}(y,y) = 1$  imply  
\begin{eqnarray}
 a_{0}(x,y|A) = \Delta^{1/2}_{VVM}(x,y) 
\label{lll} 
\end{eqnarray}
$\Delta_{VVM} $ is the  bi-scalar called Van Vleck-Morette determinant.
 It defines
the Riemannian measure in normal Riemannian 
coordinates $x^{a} \equiv x$  in ${\cal N}_{y}$ 
where $\Delta_{VVM}(x,y)^{-1} 
= \sqrt{g(x^{a})}$ 
(and thus 
$\Delta_{VVM}(y,y) = 1$)
\cite{ca}.
$\Delta_{VVM}(x,y) = - [g(x)g(y)]^{-1/2} \mbox{det}
\{\partial^2 \sigma(x,y)/\partial x^{a} \partial y^{b}\}$ holds
in general coordinates.
The mass dimensions of 
the coefficients $a_j(x,y|A)$ and the other
relevant quantities  are $[ t ] = M^{-2}$, $[ {a_{j}} (x,y|A) ] = M^{2j}$, 
$[ A ]  = M^{2}$ and
$[ K(t,x,y|A) ] = M^D $.

The expansion we have presented here (see also \cite{fu,ca,rep})
 is a bit different 
from the more usual Schwinger-De Witt
one \cite{bd,wa},
 where a further overall  exponential $\exp{-tm^2}$ appears in the right
 hand side of the expansion of $K(t,x,y|A)$,
 $m$ being the mass of the field. The  Schwinger-De Witt coefficients
 are related to those in (\ref{expansion1})  by trivial
relations \cite{ca}.
The explicit expression of some of the coefficients $a_{j}(x,y|A)$
(also for the Schwinger-DeWitt expansion and Lorentzian metrics) 
for $x=y$ and $x\neq y$ 
can be found in \cite{rep,kk} and \cite{bd} respectively.
Apart from terms depending on the particular form of $V$,
 they are polynomials in the curvature
tensors of the manifold.\\

\noindent {\bf Theorem 1.4.}
 {\em In our general hypotheses  on ${\cal M}$ and $A'$ 
 and $r>0$ we have  Weyl's formula
 \begin{eqnarray}
 \lim_{j\rightarrow +\infty}  \frac{\lambda_{j}^{D/2}}{j} = 
 \frac{(2\pi)^D}{\omega_{D} V({\cal M})}\:,  \label{weyl}
\end{eqnarray}
$\omega_{D}$ is the volume of the unit disk in $\R^D$, 
$\omega_D = \pi^{D/2}/\Gamma(1+ \: D/2)$ .}

\section{Local $\zeta$-function techniques and point-splitting procedure.}

In this section we develop the theory of the {\em local} $\zeta$ function and
then, coming back to physics,  we shall consider  how this theory 
 is employed. In particular,  proving some rigorous theorem 
 concerning the  local $\zeta$ function approach to define and 
regularize the 
physical quantities 
$\det A$ (namely $S[A]$), 
 ${\cal L}(x|A)$, $\langle \phi^{2}(x|A) \rangle $ given in {\bf 1.2}
and their relations with the corresponding point-splitting 
procedures. The case of  $\langle T_{ab}(x|A) \rangle$ will be treated in a 
 paper in preparation.

 References  concerning the physical applications 
are respectively \cite{ha,wa,bd} concerning
 the effective Lagrangian (and effective action), \cite{md}
 concerning the field fluctuations  and \cite{moa}
 concerning the averaged one-loop stress tensor.
Further references on the heat-kernel and $\zeta$-function techniques
in symmetric manifolds are \cite{ca,ze}.

\subsection{The local $\zeta$ function.}

\noindent {\bf Definition 2.1}
{\em Within our initial hypotheses on ${\cal M}$ and $A'$, and $r>0$,
the {\bf local ``off-diagonal''} $\zeta$ {\bf function} of $A$ is the function 
 defined for $Re$ $s > D/2$, $x,y \in {\cal M}$
and
$\mu^2>0$
\begin{eqnarray}
\zeta(s,x,y|A/\mu^2) := \frac{1}{\Gamma(s)}
\int_0^{+\infty} d (\mu^2 t) (\mu^2 t)^{s-1} 
\left[ K(t,x,y|A) - P_0(x,y|A) \right]\:. \label{zetad} 
\end{eqnarray}}
The mass-square parameter $\mu^2$ (almost always omitted at this step) 
is actually 
necessary from dimensional considerations and it is not fixed
from the theory as remarked above.
 $P_0(x,y|A)$ is the integral 
kernel of the  projector onto the kernel of $A$.
The given definition is well-posed (as proven within the proof of 
{\bf Theorem 2.2} below) since the integral above converges 
absolutely for $Re$ $s>D/2$ essentially because of the exponential 
decay of $K-P_{0}$ at large $t$ and the expansion (\ref{expansion1}) 
as $t\rightarrow 0^{+}$
which fixes the  bound $Re$ $s>D/2$.

All that follows (essentially based on theorems  by Minakshisundaram and
Pleijel \cite{ch})  is  a direct consequence
 of the heat kernel expansion (\ref{expansion1}),  
(\ref{sum 0}), Weyl's asymptotic formula  (\ref{weyl}) (which 
trivially implies that the series of $\lambda^{-s}$ converges for $Re$
$s>D/2$ and diverges for $Re$ $s<D/2$) )
and
the well-known identity for $a>0$, $s\in \C$, $Re$ $s>0$
\begin{eqnarray}
 a^{-s}\Gamma(s) = \int_{0}^{+\infty} dt\: t^{s-1} \: e^{-ta}\:.
\end{eqnarray}
 The properties of uniform and  absolute  convergence 
are, once again,  consequences of Mercer's theorem \cite{rn}.
In particular, the following theorem  can be proven by generalizing the 
the content of 
Remark 2 in Chapter VI, 
Section 4 of \cite{ch}. Anyhow, the first and the last  statement will be 
proven within the proof of {\bf Theorem 2.2} below.  \\

\noindent {\bf Theorem 2.1.}
{\em In our general hypotheses on ${\cal M}$ and $A'$ and $r>0$, 
for $\mu^2>0$ and $Re$ $s > D/2$,}

 (a) {\em the
integral in} (\ref{zetad}) 
{\em converges absolutely;}

(b) {\em   for $s$ fixed in the region given 
above, the function of $x$ and $y$, $\zeta(s,x,y|A/\mu^{2})$  is an 
integral kernel of the bounded  trace-class
 operator $(A/\mu^2)^{-s}$ defined by spectral theory
in the usual way, through a projector valued measure (dropping the 
spectral-measure part on the kernel of $A$ whenever it exists);}

 (c) {\em for $Re$ $s > D/2$,
  the prime meaning that any possible vanishing 
 eigenvalues and corresponding eigenvectors are omitted
from the sum,
\begin{eqnarray}
\zeta(s,x,y|A/\mu^2) = {\sum_{j\in \N }}' 
\left(\frac{\lambda_j}{\mu^{2}}\right)^{-s}
\phi_j(x)\phi_j^*(y)\:, \label{sum} 
\end{eqnarray}
where the convergence is absolute in uniform sense
 in 
$\{ s\in \C \:|\: \beta \geq  Re \mbox{ } s\geq \alpha \} \times {\cal M}\times {\cal M} $ 
for any couple $\alpha,\beta \in \R$ with $\beta >\alpha  > D/2$, and thus 
 $\zeta(s,x,y|A/\mu^{2})$ defines a 
$s$-analytic function in 
$C^0( \{ s\in \C \:|\: Re \mbox{ } s> D/2 \}
 \times {\cal M}\times{\cal M})$}.\\

We remark that $\zeta(s,x,y|A/\mu^{2})$, for $Re$ $s>D/2$, could be defined 
by (\ref{sum}) with no reference to the heat kernel, obtaining the same 
results.\\

\noindent {\bf Definition 2.2.}
{\em Within our general hypotheses on ${\cal M}$ and $A'$, and $r>0$,
the {\bf local} $\zeta$ {\bf function} of the operator $A$ is 
 the function of $x\in {\cal M}$ and $s\in \C$ with $Re$ $s>D/2$,
$\mu^{2}>0$ 
\begin{eqnarray}
\zeta(s,x|A/\mu^{2})
:= \zeta(s,x,x|A/\mu^{2}) \label{localz}\:.
\end{eqnarray}}
 {\em Similarly, the {\bf ``integrated''} 
$\zeta$ {\bf function} of $A$, 
$\zeta(s|A/\mu^{2})$ is defined by $x$ integrating
 the local one for $Re$ $s>D/2$}
\begin{eqnarray}
\zeta(s|A/\mu^2) :=  \int_{{\cal M}} d\mu_{g}(x) 
\zeta(s,x|A/\mu^{2}) \label{sum2}
\end{eqnarray}\\

Notice that  
we have from (\ref{sum}), because of the uniform convergence,
 for $Re$ $ s> D/2$,
\begin{eqnarray}
\zeta(s|A/\mu^2)  =
 {\sum_{j\in \N }}' \left(\frac{\lambda_j}{\mu^{2}}\right)^{-s} = \mbox{Tr}
 \left[ \left(\frac{A}{\mu^2}\right)^{-s}\right]\label{sum2'} 
\end{eqnarray}
The operator $A^{-s}$ is defined via spectral theory omitting
the the spectral-measure part corresponding to
 the kernel of $A$ whenever the kernel is not trivial.
As in the case of the local  $\zeta$ function, 
this series (which {\em diverges} for $Re$ $s < D/2$) converges absolutely in 
uniform sense for $s \in \{ z\in \C \:|\: \beta \geq Re \mbox{ } z \geq \alpha 
\}$, for  $\beta > \alpha > D/2$.\\

We are now able to state and prove the  most important theorem on the
local  $\zeta$
function.\\

\noindent {\bf Theorem 2.2.}
{\em Let us suppose ${\cal M}$ and $A'$ satisfy our general hypotheses 
and  also $r>0$ and 
$\mu^2 >0$}.

(a) {\em  Whenever $x\neq y$ are fixed in ${\cal M}$,}\\
(1) {\em  $\zeta(s,x,y|A/\mu^2)$  can be analytically 
continued in the whole $s$-complex plane defining an everywhere holomorphic 
function of $s$  which still satisfies} (\ref{zetad}) {\em for $Re$ 
$s>0$; moreover, this holds  everywhere in $s\in \C$ provided  
$P_{0}\equiv 0$ .} \\ 
(2) {\em The function obtained by varying $s,x,y$
belong at to  $C^{0}(\C\times (({\cal M}\times{\cal M})-{\cal D}) )$ 
together with  all of its $s$ derivative, where 
${\cal D}:= \{ (x,y)\in {\cal M}\times{\cal M}\mbox{ }|\mbox{ } x = 
y\} $.}

 (b) {\em Whenever $x=y$ are fixed in ${\cal M}$,}\\
(1) {\em    $\zeta(s,x,x|A/\mu^2)$  can be analytically 
continued in the variable $s$ into a meromorphic function 
with   possible poles, which are simple poles,
situated in the points
\begin{eqnarray}
s_j &=& D/2-j, \:\:\:\: j = 0, 1, 2,  \ldots \:\:\:\:\:\:\:\:\:\:\:\:\:\:\:\:\:\:\:
\mbox{                                 if  }  D \mbox{  is odd, or} 
\nonumber \\ 
s_j &=& D/2-j, \:\:\:\: j = 0, 1, 2, \ldots D/2 - 1 \:\:\:\:
\mbox{  if  }  D \mbox{  is even } \nonumber  
\end{eqnarray}
and residues
\begin{eqnarray}
Res (s_j) = \frac{\mu^{D-2j}a_{j}(x,x|A)}{(4\pi)^{D/2}\Gamma(D/2-j)}  
\:.
\label{residua}
\end{eqnarray}}\\
(2){\em Varying $s$ and $x$, one gets  a function 
which 
belongs  to $C^{0}((\C-{\cal P})\times {\cal M})$ 
 together  with
 all of its $s$ derivatives, where ${\cal P}$ 
  is the set of the actual poles (each for some $x$) among the 
 points listed above.}

(c) {\em For $x,y$ fixed in ${\cal M}$, the $s$-continued 
function $\zeta(s,x,y|A /\mu^{2})$
is analytic in a neighborhood of $s=0$ and
\begin{eqnarray}
\zeta(0,x,y|A/\mu^2) + P_0(x,y|A) =
\frac{a_{D/2}(x,x|A)}{(4\pi)^{D/2}}\: 
\delta_{x,y}\:\delta_{D}   \label{azero}  
\end{eqnarray}
where $\delta_{x,y}=1$ if  $x=y$ and
$\delta_{x,y}=0$ otherwise, $\delta_{D} =1$ if $D$ is even and
$\delta_{D}= 0$ if $D$ is odd.\\
  For $x\neq y$ the zero at $s=0$ of right 
hand side of} (\ref{azero}) {\em is of order $\geq 1$.}

 (d) {\em 
The analytic continuation of the integrated $\zeta$ function
coincides
    with the integral of the analytic continuation of the 
local ({\em on-diagonal}) 
 $\zeta$ function and has the same meromorphic structure of the continued
function $\zeta(s,x|A/\mu^2)$ with possible poles on the same points
and residues given by the integrals of the residues} (\ref{residua}).\\

\noindent {\em Proof.}
See {\bf Appendix} $\Box$. \\

\noindent {\em Comments.}\\
{\bf (1)}
 It is worth stressing that, whenever $A$ has no vanishing eigenvalue
(so that $P_{0}\equiv 0$) and $x\neq y$ or, equivalently, whenever  $Re$ 
$s>0$ and
$x\neq y$,
 the relation  (\ref{zetad}) 
is maintained also when the left hand side
is understood in the sense  of the analytic continuation 
and  the right hand side 
 is computed as a proper integral. This property will be very useful
 studying the Green function of $A$.\\
{\bf (2)} The simple poles of the local or integrated  $\zeta$ function are
 related to the
heat kernel coefficients in a direct way as follows from (\ref{residua}).
 It is very important to stress that there is no guarantee
that, actually, poles appear in the points indicated above because the 
corresponding coefficients $a_{j}(x,x|A)$ (or the integrated ones) may 
vanish. Anyhow, {\em if} poles appear, they must belong to the sets listed 
above. \\
{\bf (3)} As a final comment we remark that (\ref{azero}) proves that 
the continued function $\zeta(s,x,y|A/\mu^{2})$
is not continuous on the diagonal $x=y$, at least for $s=0$.
So the $s$-continuation procedure and the limit as $x\rightarrow y$
{\em generally do not commute}. \\

\noindent {\em Remark.} From now on, barring different 
specification,  the symbols of the various $\zeta$ 
functions as $\zeta(s,x,y| A/\mu^{2})$ indicate the meromorphic
 functions continued from the initial 
 domain of definition $Re$ $s>D/2$ as far as possible in the complex $s$
 plane.\\
  
  We are now able to define in a mathematical precise meaning within the 
 framework of the $\zeta$ function theory the determinant of $A$ necessary
in (\ref{integral2}).\\
 
 \noindent {\bf Definition 2.3.} 
{\em  Within our general hypotheses on ${\cal M}$ and $A'$, for $r>0$
and $\mu^{2}>0$,
the {\bf determinant} of the operator $A/\mu^{2}$ is defined as  
 \begin{eqnarray}
 \det \left( \frac{A}{\mu^{2}}\right) := e^{ 
 -\frac{d\:\:}{ds}\mid_{{s=0}} \zeta(s| A/\mu^{2})} 
 \label{detzeta}\:.
 \end{eqnarray}}

\subsection{A few comments in more general cases.}

What is it maintained of these results once one drops
 the hypotheses of a compact
manifold and/or absence of boundaries?
More general results  of the heat kernel theory
 for the pure Laplacian, with
trivial extensions to the case $A= - \nabla_a \nabla^{a} +m^2$
can be found in the literature (see \cite{wa,ch,de}).
A general discussion on the heat kernel,  considering also 
vectorial and tensorial fields and more general connections than the 
metrical one, can be found in \cite{fu}.

In 
general, the lack of the hypothesis of a compact manifolds 
 produces  the failure 
of expansions as those in (\ref{sum 0})  because the spectrum of $A$, 
the Friedrichs self-adjoint extension of $A'$
(which has to be  defined on $C_0^{\infty}(\cal{M})$ and 
still results to be essentially self-adjoint \cite{de}) becomes
continuous in general. 
One sees that, in particular cases,  
it is possible to replace the sum in (\ref{sum 0})
with integrals
dealing with opportune spectral measures
\begin{eqnarray}
K(t,x,y|A) = \int_{\sigma(A)} d\mu_{A}(\lambda) \sum_{j} \:e^{-\lambda t }
 \phi_{j\lambda}(x)
\phi_{j\lambda}^{*}(y)\:, \label{spctrl}
\end{eqnarray}
the function $\phi_{j\lambda}$ being eigenfunctions (in some  weak sense)
of $A$ with eigenvalue $\lambda$.\\
We expect that this is a general result.
This can be done also for the local
$\zeta$ function, which can be still defined by (\ref{zetad}) 
provided the corresponding integral converges.
It is anyhow  worth stressing that 
 a quite complete theory has developed in \cite{ch} for the case 
$V\equiv 0$ also considering noncompact manifolds neither
spectral measures, but thinking the non compact manifold as a limit 
 of compact (generally with boundaries) manifolds.
 In recent years, the theory of heat kernel and $\zeta$ function in
  symmetric manifolds has been developed on mathematical and physical grounds
also  proving the validity of (\ref{spctrl}) in noncompact symmetric 
 manifolds \cite{ca',rep}.
 \\
There exist  a quite large theoretical-physics literature on 
applications of these 
topics \cite{rep} in quantum field theory in curved background,
 concerning particular cases and also higher spin of the field
(see \cite{ielmo,md} for the case of photons and gravitons 
spaces containing conical singularities), in particular, in the
 presence of  noncompact manifolds  containing conical singularities
very important within QFT in the presence of a black hole 
\cite{articoloZERBINIcognolavanzo}. (Problems related to the heat kernel 
and the $\zeta$ function in the case 
 of a compact manifold containing
conical singularities is not trivial on a mathematical point of view,
it was treated by J. Cheeger in \cite{cheeger}.).
It is known that, in the case $V\equiv m^{2}$ at least,
 both the regularity (including the positivity)
  of $ K(t,x,y)$ and the heat kernel expansion 
(\ref{expansion1}) do not depend on the compactness of the
manifold provided further hypotheses on ${\cal M}$ and $A$ are given
 (see \cite{wa,ch,de}). 
In particular, the behavior at 
$t\rightarrow 0^+$ is the same as in
 the compact case and one finds the asymptotic
expansion (\ref{expansion1}) once again \cite{wa}.
Generally speaking, provided $K(t,x,y)$ is given through a spectral 
measure and $A$ is  positive definite, one can still prove, on any 
compact $K \subset{\cal M}$, 
(\ref{magg}) where now $\lambda = \inf \sigma(A)$. In this way
$\zeta(s,x,y|A/\mu^{2})$ can be defined and  the 
results of {\bf Theorem 2.2} remain substantially unchanged.
 General estimates on $(x,y)$-uniform bounds
 of $K(t,x,y|A)$ at large and little $t$ can be found in \cite{ch,de} 
 for the pure Laplacian
 imposing further requirement on the geometry of the manifolds and
 bounds on the Ricci operator and sectional curvatures.
 The presence of  vanishing (proper) eigenvalues can be treated similarly to 
 the case of compact manifolds, subtracting 
  the 
 contribution of the corresponding eigenvectors from the heat kernel
 as given in (\ref{zetad}).  
 The existence of the integrated 
$\zeta$ function is much more difficult to study in the general case of a 
noncompact manifold, 
and, barring very particular situations (e.g. the Euclidean section
of  anti de Sitter spacetime), the integrated $\zeta$ function
does not exist  and one needs some volume cutoffs. Recently, other ways 
to overcome this shortcoming has been pointed out \cite{mu}.
The problem of the existence of the integrated $\zeta$ function  is 
dramatically important in the issue of the computation of 
thermodynamical  quantities of fields propagating in the spacetime 
around a black hole (and thus in the general issue of the black hole 
entropy). In that case also horizon divergences appear and their r\^{o}le 
and involved  mathematics is not completely understood also because 
different integrated renormalization procedures disagree 
\cite{articoloZERBINIcognolavanzo,ielmo,md,mob,ff,ie}.
Finally, the issue whether or not $-(2\beta)^{-1} \zeta'(0,x|A/\mu^{2})$
defines the true local density of free energy still remains an open question
 (see the discussion in \cite{moa}).
The presence of boundaries, maintaining the compactness,  changes
 the results obtained in the case above ("closed" manifold)
  only for the presence of further 
terms in the heat-kernel expansion (\ref{expansion1}) depending on the 
boundaries \cite{ch} and with noninteger powers of $t$. These terms 
can be interpreted as distributions concentrated on the boundary of the 
manifold (see \cite{ze,el} for the corresponding bibliography and physical
 applications).
Obviously, in this case 
 $A'$ is not essentially self-adjoint and the considered self-adjoint
extension depends on the imposed boundary conditions (Dirichlet/Neumann/Mixed 
problem) and the choice is related 
to the particular quantum state one is investigating.

\subsection{$S[A]$ and ${\cal L} (x|A)$.}
 
 The results of these two  subsections are quite known 
 \cite{wa,bd,fu} in particular cases (e.g. $D=4$, $V= m^{2} + 
 \xi R$, $m^{2}>0$) and, barring \cite{wa}, just in a formal way. Herein,
 we
  produce a rigorous proof of the substantial
 equivalence  of the point-splitting approach and $\zeta$ function
 procedure as far  effective Lagrangian is concerned, starting 
 from our hypotheses\footnote{These
  hypotheses and our way are different from those used in \cite{wa}
  which considered the case $V\equiv m^{2 }$ only.
   There (also dropping the hypotheses of a compact manifold)   
   the Schwinger-DeWitt expansion  and
    an explicit use of large $t$ behaviour of ($t$-derivatives 
  of) $K(t,x,y)$ were used together 
  with, in part, hypotheses of the essentially self-adjointness of $A'^n$ 
  for any $n\in \N$. The procedures used in the fundamental book \cite{bd}
   are very formal and no mathematical discussion appear.} 
  in  the general case $D >1 $.   
 We shall give also  some 
 comments on some formal
 definitions often  assumed by physicists.  
 We remark also that, 
 the relation between the  local $\zeta$ function approach and
 the point-splitting procedure is now discussed
 in terms of the expansion (\ref{sum 0}) instead of the Schwinger-DeWitt
 one (hence, differently from other papers on the same subject,
   the obtained  formulae do not distinguish between 
 the cases of a finite or vanishing mass of the field)
 and a quite  general scalar operator is considered here.
In favour of Schwinger-DeWitt's expansion,
 it might be noticed anyhow that this expansion, at least formally
and for $m$ strictly positive, should work also in noncompact manifolds
due to the sharp decay of the exponential $\exp (-m^2 t)$ in the heat 
kernel (see discussion in \cite{wa}).\\
  
 \noindent {\bf Definition 2.4.}
 {\em Within the general hypotheses on ${\cal M}$ and $A'$ and $r>0$,
 the {\bf effective action} associated to the operator $A$, is  defined,
 within the $\zeta$ function approach, as
 \begin{eqnarray}
 S[A]_{\mu^{2}} := - \ln Z[A]_{\mu^{2}}  \label{ea}\:,
 \end{eqnarray}
 where the {\bf partition function} $Z[A]_{\mu^{2}}$ in the right hand side 
 is defined as}
 \begin{eqnarray}
Z[A]_{\mu^{2}} := \left[ \mbox{det} \left( \frac{A}{\mu^{2}}\right) 
\right]^{-1/2} =
e^{\frac{1}{2}\frac{d\:\:}{ds}|_{s=0}\zeta(s|A/\mu^{2})} 
\label{zaza} \:.
\end{eqnarray}  

\noindent Therefore we have   defined $Z[A]$
 by (\ref{integral2}), the left hand side  being 
rigorously interpreted as in (\ref{detzeta}) in the framework of the 
$\zeta$-function.
 The definition of the effective Lagrangian is similar.
 The most natural choice is the following definition 
 (where from now on the 
 prime on a $\zeta$ function means the $s$ derivative)\\
 
 \noindent {\bf Definition 2.5.}
 {\em Within the general hypotheses on ${\cal M}$ and $A'$ and $r>0$,
 the {\bf effective Lagrangian} associated to the operator $A$, is  defined,
 within the $\zeta$ function approach, by}
  \begin{eqnarray}
 {\cal L}(x|A)_{\mu^{2}}  := -\frac{1}{2}\zeta'(0,x|A/\mu^{2})  \:. 
 \label{effaction}
\end{eqnarray}

Notice that (\ref{effective})
 is now fulfilled by definition of integrated
$\zeta$ function (\ref{sum2}).
Furthermore, it is worth stressing  that definition 
(\ref{effaction}) is well-posed
because {\bf Theorem 2.2} states that no singularity can apper at
$s=0$ in the local $\zeta$ function.

We want to comment this definition to point out what such a definition
 actually "regularizes". This is also to make precise what is actually 
 allowed  and what is forbidden within the $\zeta$ function approach. 
 Following \cite{wa} and starting from (\ref{ea}), 
 using (\ref{zaza}) and (\ref{detzeta}), one has correctly (we omit 
 the index $\mu^{2}$ for sake of simplicity in the notations)
 \begin{eqnarray}
 	   S[A] = -\ln Z[A] =    
 	  -\ln  \left[\mbox{det}\left(\frac{A}{\mu^{2}} \right)\right]^{-1/2}.
 	  \label{correct}
 \end{eqnarray}
 At this step and quite often, physicists  assume   the  
 validity of the matrix identity (for the moment we omit the 
 factor $\mu^{2}$ which is not necessary)
 \begin{eqnarray}
 \mbox{Tr} \ln A = \ln \det A \:. 
\label{wrong}
 \end{eqnarray}
 One may define at this end (in the strong operatorial  topology)
\begin{eqnarray}
\ln A = \lim_{\epsilon \rightarrow 0^{+}} \left\{
 \int_{\epsilon}^{+\infty} dt \: \frac{e^{-t A}}{t}
 + (\gamma - \ln \epsilon) I  
\right\}\:.
\label{ln}
\end{eqnarray} 
 Anyhow,  the trace of this operator {\em does not exist} at 
  least because of the presence of $\gamma I$. 
    Also a direct definition
  of $\ln A$ by spectral theorem would prove that $\ln A$ is not
  a bounded operator and thus, {\em a fortiori}, it is not a  trace-class
   operator. We conclude that (\ref{wrong}) {\em makes no sense}
  in any cases, neither within the $\zeta$-function approach. 
 Anyhow, it is still possible to grasp our definition by this way 
 employing a completely formal sequence of identities.  
 The way is just to  drop the annoying term in (\ref{ln}) as well as the 
  regulator $\epsilon$ and write down through (\ref{wrong}),
  using (\ref{trace}), (\ref{zetad}) (dropping $P_{0}(x,x)$ for sake 
  of simplicity)
 and interchanging several times  the order of trace symbol 
 and integrals and the $s$-derivative
  \begin{eqnarray}
 -2 S[A] &=&
 \mbox{Tr} \ln A = 
 \int_{0}^{+\infty} \frac{dt}{t} \: \mbox{Tr}\: e^{-t A}  =
 \int_{0}^{+\infty} \frac{dt}{t} \: \mbox{Tr}\: K_{t}   \nonumber\\
&= &  \int_{{\cal M}} 
\:d\mu_{g}(x)\: 
\int_{0}^{+\infty} \:dt\: t^{-1} K(t,x,x|A) \nonumber \\ 
&=&  
 \int_{{\cal M}} 
\:d\mu_{g}(x)\: 
\int_{0}^{+\infty} \:dt\: \frac{d\:\:\:}{ds}|_{s=0}\left(
\frac{1}{\Gamma(s)} t^{s-1}\right) K(t,x,x|A) \nonumber \\
&=&  \int_{{\cal M}} 
\:d\mu_{g}(x)\: \zeta'(0,x|A)    \label{f} \:.
\end{eqnarray} 
Notice that, above, also the $t$-integration of the heat kernel 
times $t^{-1}$ evaluated 
for $x=x$ trivially diverges, and thus also the first passages 
 above are incorrect.
   Anyhow,
looking at the last side, 
it is natural, from the formal identities above
to get the definition (\ref{effaction}).
${\cal L}(x|A)$ may be hence  considered as the "formal" integral kernel of the 
operator $\ln A$ evaluated on the diagonal.

As a final comment on (\ref{wrong})  we remark that, nevertheless, 
one could use this
(literally wrong) identity to {\em define} an extension  of the concept of the 
trace of $\ln A$, this is because the right hand side is however
 well defined.
Anyhow, this way leads to a generally {\em not linear} trace due to the 
well-know {\em multiplicative anomaly} of the determinant defined in 
terms of  $\zeta$ function  \cite{manomaly}.

\subsection{Effective action and point-splitting procedure.}

Let us consider the relation between the employed definitions and the
point-splitting renormalization procedure. The idea of the 
point-splitting procedure  \cite{wa,bd} consists, following the formal 
passages developed in (\ref{f}), of  the formal definition 
for the
effective Lagrangian  
\begin{eqnarray}
 {\cal L}(y|A) :=  \lim_{x\rightarrow y} 
\left [-\int_{0}^{+\infty} \:\frac{dt}{2t}\:  K(t,x,y|A) - 
 \mbox{      "divergences"    } \right] \label{pointL}.
\end{eqnarray}
The divergences are those which appear evaluating 
the 
limit in the integral  above by brute force 
\cite{wa,bd}. We shall 
find the precise form of these by our local $\zeta$ function 
approach. We notice that the term "divergences" is quite ambiguous, because
a divergent term plus a finite term is always a divergent term.
Such an ambiguity could be  actually  expected \cite{wa} because of
 dependence on $\mu$ at least, which has to remain into the final expression 
of the renormalized effective action for several reasons 
\cite{wa,ha,bd} at least for $D=4$.
Therefore  the final expression should  
contain a finite renormalization part dependent on the arbitrary scale
$\mu$. In practice, the actual value of $\mu$ can be fixed by 
experimental results (at least for $D=4$).
As a further general comment, which is not so often remarked,
 we stress that the 
point-splitting procedure  {\em works only in the 
case} $P_{0}(x,y)\equiv 0$. Indeed, whenever $\lambda = 0$ 
is an eigenvalue of 
$A$, the integral for $x\neq y$ in (\ref{pointL}) diverges due to the 
large $t$ behaviour of the integrand. To avoid this drawback,
one could try to change the integrand  into $K(t,x,y|A) - P_{0}(x,y)$.
Nevertheless this is not enough, indeed, a straightforward check at  
$t\rightarrow 0^{+}$ using {\bf Theorem 1.3}, proves 
that the integrand so changed inserted  in (\ref{pointL})
 produces a divergent integral at $t\rightarrow 0^{+}$ just because of the 
 presence of $P_{0}$. 

We need a definition and a useful lemma to prove that, within our general 
hypotheses and supposing also $r>0$, the previous subtraction of 
divergences together with the coincidence limit
are actually equivalent to the definition given in the local $\zeta$ function 
framework (\ref{effaction}).\\

\noindent {\bf Definition 2.6.}
{\em Within our general hypotheses on ${\cal M}$ and $A'$ and $r>0$, for $N$ 
integer $>D/2 +2$  and $\mu^2, \mu_0^2  >0$ fixed,  
the $N$ {\bf truncated local} $\zeta$ {\bf function}
is defined as the function of $s\in \C$, $x,y\in {\cal M}$
where the right hand side makes sense
\begin{eqnarray}
\zeta(N,s,x,y| A/\mu^{2}, \mu_0^{-2}) &:=&
  \zeta(s,x,y|A/\mu^{2})
 + \left(\frac{\mu}{\mu_{0}} \right)^{2s}
 \frac{P_{0}(x,y)}{s\Gamma(s)}
 \nonumber\\
& &  -\frac{\mu^{2s}\chi (\sigma(x,y))}{(4\pi)^{D/2}\Gamma(s)}
 \sum_{j=0}^N a_j(x,y|A)\int_{0}^{\mu_{0}^{-2}} \frac{dt}{t} t^{s+j-D/2}
{e^{-\frac{\sigma(x,y)}{2t}}}
  \label{dim''}
 \end{eqnarray}}

Several properties of the function defined above are analized in the proof 
of {\bf Theorem 2.2} given in {\bf Appendix}.\\

\noindent {\bf Lemma 2.1.} {\em Within our general hypotheses on ${\cal 
M}$, $A'$ and for $r>0$, $\mu,\mu_{0}>0$ and $N>D/2 +2$,
 the function  $(s,x,y)\mapsto \zeta(N,s,x,y|A/\mu^{2}, \mu_{0}^{2})$
 is  analytic  in  $\{ s\in \C \mbox{ } | \mbox{ } Re\mbox{ } s> 
D/2 - N\}$ and  
 belongs to  $C^{0}( \{ s\in \C \mbox{ } | \mbox{ } Re\mbox{ } s> 
D/2 - N\}
 \times{\cal M}\times{\cal M})$ together with all of its $s$ derivatives.
 Moreover}
\begin{eqnarray}
\zeta'(N,0,x,y|A/\mu^{2}, \mu_{0}^{-2}) =
\Gamma(s) \zeta(N,s,x,y|A/\mu^{2}, \mu_{0}^{-2})|_{s=0} \label{KUKU}
\end{eqnarray}
{\em Finally, 
\begin{eqnarray}
\zeta(s,x| A/\mu^{2}) &=& \zeta(N,s,x,x|A/\mu^{2}, \mu_{0}^{-2}) 
 -\frac{(\mu/\mu_{0})^{2s} P_{0}(x,x)}{s\Gamma(s)} \nonumber\\
 & & +  \frac{\mu^{2s}}{(4\pi)^{D/2}}
\sum_{j=0}^{N}
\frac{a_{j}(x,x|A) (\mu_{0}^{-2})^{(s+j-D/2)}}{\Gamma (s) (s+j-D/2)}
\label{dim2''}\:.
\end{eqnarray}}

\noindent{\em Proof.} The first part and the last identity are proven 
within the proof of {\bf Theorem 2.2}. (\ref{KUKU})  is a 
trivial consequence of (\ref{s}) in the proof of {\bf Theorem 2.2}
noticing that, there,  the derivative
at $s=0$ of the analytic continuation of $1/\Gamma(s)$ is equal to $1$
and  $1/\Gamma(s)\rightarrow 0$ as $s\rightarrow 0$. 
$\Box$\\

Now, let us consider  the identity (\ref{dim2''}) ($N>D/2+2$) in the case 
$P_{0}\equiv 0$ and evaluate the $s$ derivative for $s=0$ necessary to 
get ${\cal L}(x|A)$ by (\ref{effaction}). We have 
\begin{eqnarray}
\zeta'(0,y| A/\mu^{2}) & =& \zeta'(N,0,y,y|A/\mu^{2}, \mu_{0}^{-2}) 
 -\sum_{j=0,j\neq D/2}^{N} \frac{\mu_{0}^{(D-2j)}a_{j}(y,y|A)}{ (4\pi)^{D/2}
 (D/2-j)}
\nonumber\\
& & + \delta_{D}\:\left[\gamma + \ln \left( \frac{\mu}{\mu_{0}}
\right)^{2} \right] \frac{a_{D/2}(y,y|A)}{(4\pi)^{D/2}} \label{ququ}\:,
\end{eqnarray}
where as usual, $\delta_{D}=1$ if $D$ is even  and  $\delta_{D}=0$ if
$D$ is odd. From {\bf Lemma 2.1}, we get 
\begin{eqnarray}
 \zeta'(N,0,y,y|A/\mu^{2}, \mu_{0}^{-2}) &=& \lim_{x\rightarrow y}
\left[ \Gamma(s) \zeta(N,s,x,y|A/\mu^{2}, \mu_{0}^{-2})|_{s=0} 
\right]\nonumber\:. 
\end{eqnarray}
By (\ref{dim''}), taking also account of $P_{0}\equiv 0 $ in (\ref{zetad}), 
the right hand side of the equation above reads
\begin{eqnarray}
 & & \lim_{x\rightarrow y} \left[ 
\Gamma(s) \zeta(s,x,y|A/\mu^{2})|_{s=0}
-\sum_{j=0}^N \frac{a_j(x,y|A)}{(4\pi)^{D/2}}
\int_{0}^{\mu_{0}^{-2}} dt\: t^{j-D/2-1} 
{e^{-\sigma(x,y)/2t}} \right] \nonumber \\
 &=&  \lim_{x\rightarrow y} \left[ 
\int_0^{+\infty} \frac{dt}{t}\:  K(t,x,y|A) 
-\sum_{j=0}^N \frac{a_j(x,y|A)}{(4\pi)^{D/2}}
\int_{0}^{\mu_{0}^{-2}} dt\: t^{j-D/2-1} 
{e^{-\sigma(x,y)/2t}} \right] \label{monster}
\end{eqnarray}
Notice that the integral of the heat-kernel times $t^{-1}$ converges
away from the diagonal as follows from the proof of {\bf Theorem
2.2} (see {\bf Appendix}).
To conclude,  we have just to substitute the obtained result into 
the right hand side of (\ref{ququ}) and using  definition (\ref{effaction})
and finally we have
\begin{eqnarray}
{\cal L}(y|A) &=&  \lim_{x\rightarrow y} \left\{ 
-\int_0^{+\infty} \frac{dt}{2t}\:  K(t,x,y|A) 
+\sum_{j=0}^N \frac{ (\frac{\sigma}{2})^{j-D/2}
a_j(x,y|A)}{2(4\pi)^{D/2}}
\int_{\frac{\sigma\mu_{0}^{2}}{2}}^{+\infty}
 du\: u^{D/2-j-1} 
{e^{-u}} \right. \nonumber \\
& & \left. +\sum_{j=0,j\neq D/2}^{N} \frac{\mu_{0}^{(D-2j)} a_{j}(y,y|A)}{2
(4\pi)^{D/2}(D/2-j)} - \delta_{D}\left[\gamma + \ln \left( \frac{\mu}{\mu_{0}}
\right)^{2} \right]
\frac{a_{D/2}(y,y|A)}{2(4\pi)^{D/2}}\right\} \label{monster2}\:.
\end{eqnarray}
It is possible to compute more explicitly the integrals above and give 
a close form of  the divergent terms in (\ref{monster2}).
This can be done expanding the integrals in powers/logarithm of $\sigma$
 and keeping both the dominant divergent terms and those constant only.
 In the following  $j = 0, 1, 2, \ldots D/2+1, \ldots N$ ($N \geq 
 D/2+2$) and  $D>1$.
 Let us define, with a little abuse of notation since 
 the right hand side is not function of $\sigma$ only, 
 \begin{eqnarray}
 H_{j}(\sigma) := \frac{ (\frac{\sigma}{2})^{j-D/2}
a_j(x,y|A)}{2(4\pi)^{D/2}}
\int_{\frac{\sigma\mu_{0}^{2}}{2}}^{+\infty}
 du\: u^{D/2-j-1} 
{e^{-u}}\:. \label{HJ}
\end{eqnarray}
Therefore one gets by some computations,
 $O_{k}(\sigma)$ being functions 
 which vanish as $x \rightarrow y$, in the case of $D$ even
\begin{eqnarray}
H_{j \geq D/2+1}(\sigma) &=& - \frac{\mu_{0}^{(D-2j)}}{D/2-j}
\frac{a_{j}(y,y|A)}{2(4\pi)^{D/2}} +  O_{j}(\sigma)\:; \label{pri}\\
H_{D/2}(\sigma) &=& - \frac{a_{D/2}(x,y|A)}{2(4\pi)^{D/2}} \ln \left(
\frac{\sigma \mu_{0}^{2}}{2} \right) - \gamma 
\frac{a_{D/2}(y,y|A)}{2(4\pi)^{D/2}} + O_{D/2}(\sigma)\:;\\
H_{j<D/2}(\sigma) &=& (D/2-j-1)!\frac{a_{j}(x,y|A)}{2(4\pi)^{D/2}} \left(
\frac{2}{\sigma} \right)^{D/2-j} 
- \frac{\mu_{0}^{(D-2j)}}{D/2-j}
\frac{a_{j}(y,y|A)}{2(4\pi)^{D/2}} 
+ O_{j}(\sigma)
\end{eqnarray}
and, whenever $D$ is odd, 
\begin{eqnarray}
H_{j>(D+1)/2}(\sigma) &=& - \frac{\mu_{0}^{(D-2j)}}{D/2-j}
\frac{a_{j}(y,y|A)}{2(4\pi)^{D/2}} + O_{j}(\sigma) \:; \\
H_{(D+1)/2}(\sigma) &=&\frac{a_{(D+1)/2}(y,y|A)}{\mu_{0}(4\pi)^{D/2}}
+ O_{(D+1)/2}(\sigma) \:; \\
H_{(D-1)/2}(\sigma) &=&  \frac{a_{(D-1)/2}(x,y|A)}{2(4\pi)^{D/2}} 
\sqrt{\frac{2\pi}{\sigma}} - \frac{\mu_{0}a_{(D-1)/2}(y,y|A)}{(4\pi)^{D/2}}
 + O_{(D-1)/2}(\sigma)\:;\\
H_{j<(D-1)/2}(\sigma) &=& \left( \frac{2}{\sigma}\right)^{D/2-j}
\frac{(D-2-2j)!! \sqrt{\pi}}{2^{(D+1)/2-j}} \frac{a_{j}(x,y|A)}{(4\pi)^{D/2}}
-\frac{\mu_{0}^{(D-2j)}}{D/2-j}\frac{a_{j}(y,y|A)}{2(4\pi)^{D/2}}\nonumber\\  
& & +  O_{j}(\sigma)  \label{ul}\:.
\end{eqnarray}
By substituting the results above into (\ref{monster2}), we have\\

\noindent {\bf Theorem 2.3} {\em Within our general hypotheses on 
${\cal M}$ and $A'$, for
$r>0$ and $\mu>0$ and $D>1$, the effective action computed by} 
(\ref{effaction})
{\em can be also computed by a point-splitting procedure provided 
$P_{0}\equiv 0$. Indeed,  whenever $D$ is even}
\begin{eqnarray}
{\cal L}(y|A)_{\mu^{2}} &=& \lim_{x\rightarrow y} \left\{
-\int_{0}^{+\infty}\frac{dt}{2t}\:K(t,x,y|A) -
\frac{a_{D/2}(x,y|A)}{2(4\pi)^{D/2}} \ln \left( \frac{\sigma \mu^{2}}{2}\right)
\right. \nonumber\\
 & &+ \left.
 \sum_{j=0}^{D/2-1}(D/2-j-1)!\frac{a_{j}(x,y|A)}{2(4\pi)^{D/2}}   
 \left( \frac{2}{\sigma}\right)^{D/2-j}\right\} - 2\gamma 
 \frac{a_{D/2}(y,y|A)}{2(4\pi)^{D/2}} \:, \label{even} 
\end{eqnarray}
 {\em and,  whenever  $D$ is odd}.
\begin{eqnarray}
{\cal L}(y|A)_{\mu^{2}}
 &=& \lim_{x\rightarrow y} \left\{
-\int_{0}^{+\infty}\frac{dt}{2t}\:K(t,x,y|A) +
\sqrt{\frac{2\pi}{\sigma}}\frac{a_{(D-1)/2}(x,y|A)}{2(4\pi)^{D/2}} 
\right. \nonumber\\
 & &+ \left.
 \sum_{j=0}^{(D-3)/2}\frac{(D-2j-2)!! \sqrt{\pi}}{2^{(D+1)/2-j}}
 \frac{a_{j}(x,y|A)}{(4\pi)^{D/2}}   
 \left( \frac{2}{\sigma}\right)^{D/2-j}\right\}\:. \label{odd} 
\end{eqnarray}

\noindent {\em Comments.}\\
{\bf (1)}  When $D$ is odd,  $\mu$ disappears from the final results. 
Conversely, when $D$ is $even$,
 the scale $\mu$ appears, and this  is necessary due 
to the logarithmic divergence in (\ref{even}), indeed, it has to combine 
with $\sigma$ in order to give a nondimensional argument of the 
logarithm. Since the presence of $\mu$ in (\ref{even}), the left hand 
side is ambiguously defined because it can be changed by adding terms
of the form, where $\alpha$ is any strictly positive real,
\begin{eqnarray}
\delta {\cal L}(x,\alpha |A) :=
 -  \delta_{D} \frac{a_{D/2}(x,x|A)} {2(4\pi)^{D/2}}  \ln \alpha \:,
\end{eqnarray}
This correspond trivially to a rescaling of $\mu^{2}$: $\mu^{2 } 
\rightarrow \alpha \mu^{2}$.
These terms cannot be determined within this theory and represent a 
remaining finite part of the renormalization procedure.
The pointed out ambiguity is a subcase of an ambiguity which arises 
 also in the presence of $P_{0}$.
 This can be carried out directly from (\ref{effaction}).
In fact, directly from 
the definitions (\ref{localz}) and (\ref{zetad}), we have
$\zeta(s,x| A/(\alpha \mu^{2})) = \alpha^s \zeta(s,x| A/\mu^{2})$
and thus 
$\zeta'(0,x| A/(\alpha \mu^{2})) = \zeta'(0,x| A/\mu^{2})
+ \zeta(0,x| A/\mu^{2}) \ln \alpha$.
Reminding (\ref{azero}) and (\ref{effaction}) we get 
\begin{eqnarray}
 {\cal L }(x|A)_{\alpha\mu^{2}} = {\cal L }(x|A)_{\mu^{2}} 
 - \left[ \delta_{D} \frac{a_{D/2}(x,x|A)}{2(4\pi)^{D/2}} 
 -\frac{P_{0}(x,x)}{2}\right] \ln \alpha\:. 
\end{eqnarray}
{\bf (2)} All these results should remain unchanged also in the case of a 
noncompact manifold because all proofs was based on {\bf Theorem 2.2},
which, as discussed in {\bf 2.2} should hold true also
dropping the hypothesis of compactness (and assuming some further hypotheses
as completeness).

\subsection{ $\langle \phi^2(x|A) \rangle $ and local $\zeta$ function.}

Let us consider the local $\zeta$-function definition of the field
fluctuations  $\langle \phi^{2}(x|A)\rangle$
 \cite{dm}.  
The main  definitions \cite{dm} are the following ones\\

\noindent {\bf Definition 2.7.} 
{\em Within our general hypotheses on ${\cal M}$ and $A'$ and $r>0$, the 
{\bf field fluctuation} of the field associated to the operator $A$
are defined by
\begin{eqnarray}
\langle \phi^{2}(x|A)\rangle_{\mu^{2}}
 := \frac{d\:\:}{ds}|_{s=0} Z(s,x| A/\mu^{2})
\label{phiz}\:,
\end{eqnarray}
where the {\bf local} $\zeta$ {\bf function of the field fluctuations}
 $Z(s,x| A/\mu^{2})$ is 
defined as the function of $x$ and $s$ whenever the right hand side 
is sensible, for any $\mu^{2}>0$
\begin{eqnarray}
Z(s,x| A/\mu^{2}) := \frac{s}{\mu^{2}} \zeta(s+1,x| A/\mu^{2}) \:.
\label{phiz'}
\end{eqnarray}}

Concerning  the mathematical 
consistency of the  proposed definitions,
from {\bf Theorem 2.2}, we have \\

\noindent {\bf Theorem 2.4.}
{\em In our hypotheses on ${\cal M}$ and $A'$ and $r>0$, for $\mu^2 >0$,
  $(s,x) \mapsto Z(s,x|A/\mu^2)$  is a  meromorphic function of $s$ 
analytic in $s=0$
and   
the only  possible poles  are {\em simple} poles 
and are
situated in the points
\begin{eqnarray}
s_j &=& D/2 - j - 1, \:\:\:\:\: j = 0, 1, 2,  \ldots \:\:\:\:\:\:\:\:
\mbox{     if  }  D \mbox{  is odd, or} \nonumber \\ 
s_j &=& D/2-j -1, \:\:\:\: j = 0, 1, 2, \ldots D/2 - 2 \:\:\:\:
\mbox{  if  }  D \mbox{  is even. } \nonumber 
\end{eqnarray} 
Moreover the function $Z$ and all of its $s$ derivatives belong
to $C^{0}((\C-{\cal P})\times {\cal M})$, ${\cal P}$ being the set of the 
actual poles (each for some values of $x$) among the points  listed above.}\\

\noindent {\em Comments.}\\
{\bf (1)} Let us summarize the formal procedure which 
leads one to the definitions above \cite{dm}.
 The general idea consists of considering  the following 
 {\em purely formal} identity
which takes account of the formal definition  (\ref{formal}) and
the rigorous identity  (\ref{sum2})
\begin{eqnarray}
 \langle \phi^{2}(x|A) \rangle_{\mu^{2}} &=& -\frac{\delta\:\:}{\delta J(x)}
 |_{J\equiv0}
 \frac{1}{2}\frac{d\:\:}{ds}|_{s=0}
 {\sum_{j\in \N}}'
 \left\{ \frac{\lambda_{j}[A+2J]}{\mu^{2}}\right\}^{-s}\nonumber \\
 &=& 
 - \frac{d\:\:}{ds}|_{s=0}{\sum_{j\in \N}}' 
 \frac{\delta\:\:}{\delta J(x)}|_{J\equiv 0}
 \left\{
  \frac{\lambda_{j}[A+J]}{\mu^{2}}\right\}^{-s} \label{inter}\:.
 \end{eqnarray}
Then one formally computes the functional derivatives (G\^{a}teaux 
derivatives) of 
$\{\lambda_{j}[A+J]/\mu^{2 }\}^{-s}$ at $J\equiv 0$ 
\cite{dm,moa} obtaining
\begin{eqnarray}
\frac{\delta\:\:}{\delta J(x)}|_{J\equiv 0}
 \left\{ \frac{\lambda_{j}[A+J]}{\mu^{2}}\right\}^{-s} =  -\frac{s}{\mu^2} 
 \left[\frac{\lambda_{j}}{\mu^{2}}\right]^{-(s+1)}
  \phi_{j}(x)\phi^{*}_{j}(x) \sqrt{g(x)}.  
\end{eqnarray}
This result, inserted in (\ref{inter}) and interpreting the final series in 
the sense of the analytic continuation, gives both (\ref{phiz}) and
 and (\ref{phiz'}). Obviously one could try to give some rigorous meaning
to the formal passages above, but this is not our approach, which 
assumes
  (\ref{phiz}) and (\ref{phiz'}) by {\em definition}.\\
{\bf (2)} In the case $\zeta(s,x|A/\mu^{2})$ has no pole at $s=1$,
namely, when $D$ is odd or when $D$ is even and $a_{D/2-1}(x,x) = 0$
(see {\bf Theorem 2.2}) (\ref{phiz}) reduces to the trivial formula, which 
does not
depend of the value of $\mu^{2}$ (this follows directly from 
the definition of local $\zeta$ function (\ref{localz}))
\begin{eqnarray}
\langle \phi^{2}(x|A)\rangle_{\mu^{2 }}
= \mu^{-2} \zeta(1,x|A/\mu^{2}) = \zeta(1,x|A)\:. \label{simple} 
\end{eqnarray}
{\bf (3)} We finally remark that, in \cite{dm}, 
 definitions (\ref{phiz}) and (\ref{phiz'}) have been 
 checked  
 on several concrete cases obtaining a perfect agreement with other 
 renormalization procedures, also concerning the remaining finite 
 renormalization part related to the $\mu^{2}$ ambiguity.
 The local $\zeta$-function approach concerning the field fluctuations
   has  produced also a few  new 
 results, e.g., the general form for renormalized 
 trace of the one-loop stress tensor in the generally
 nonanomalous case, and several applications in symmetric spaces for 
 general values of the parameter $\xi$ which fixes the coupling of
 the field with the curvature (see \cite{dm}).

 \subsection{$\mu^{-2n}\zeta(n,x,y|A/\mu^{2})$ as Green function
  of $A^n$.}
 
  Let us consider the usual  operator $A$, the Friedrichs extension of 
  $A'$ given in (\ref{d}).  Let us also  suppose explicitly that
 $P_{0}   \equiv 0$ namely, 
$\sigma(A) \subset (0, +\infty)$.
 In this case
  $A$ has a well-defined  unique inverse operator  $A^{-1}$ : 
$R(A)$ 
$\rightarrow$   ${\cal D}(A)$. 
We notice that $A^{-1}$ is bounded by
 $\sup \left\{  1/\lambda | \lambda \in \sigma(A)\right\} < +\infty$. 
Moreover
   $R(A)$ is dense in $L^{2}({\cal M}, d\mu_{g})$, 
   because  $R(A) = R(A-0I)$ which is dense $A$ being self-adjoint 
   and  $0$ belonging to the 
   resolvent $\rho (A)$. Therefore $A^{-1}$ can be uniquely 
   extended into a bounded  
   operator defined on the whole $L^{2}({\cal M}, d\mu_{g})$ which we 
   shall indicate with the same symbol $A^{-1}$.
    (Alternatively, one can check on the fact that 
   $R(A)$ is dense in $L^{2}({\cal M}, d\mu_{g})$ directly from the
   spectral representation of $A$, where the series is understood 
   in the strong topology,
  \begin{eqnarray}
  A = \sum_{j=0}^{+\infty} \lambda_{j} \phi_{j}(\phi_{j},\:\:\: )\:,
  \end{eqnarray}
  taking account that the vectors $\phi_{j}$ defines a Hilbertian 
  base of $L^{2}({\cal M}, d\mu_{g})$).
  By definition of inverse operator,
  $A A^{-1} = I$ holds true 
  in the whole space and not only in  $R(A)$, being 
   $A^{-1}(L^{2}({\cal M},d\mu_{g}))  \subset {\cal D}(A)$ (this follows  
 from the fact that   $A=A^{\dagger}$ is  a closed 
  operator and $A^{-1}$ is bounded in the dense 
  set $R(A)$). The other relation $A^{-1} A = I_{{\cal D}(A)}$ holds 
  true in ${\cal D}(A)$ as indicated by the employed notation.\\
   We are now interested in integral 
  representations of $A^{-1}$.
  In the case $D < 4$, one gets from {\bf Theorem 1.4} that the 
  series of elements $ || A^{-1}\phi_{j}||^{2} 
  = |\lambda_{j}|^{-2} = \lambda_{j}^{-2}$
   converges and, since $\left\{ \phi_{j}| j= 0,1,2, \ldots \right\}$
   is a Hilbertian base,
   this means that \cite{rs}
  $A^{-1}$  is a Hilbert-Schmidt operator and thus, holding our 
  hypotheses of a countable topology involving a consequent separable measure, 
   $A^{-1}$ is represented by
  an $L^{2}({\cal M}\times{\cal M})$ integral kernel $A^{-1}(x,y)$. 
  As is well known, 
  such a  function is called a {\em Green function} of $A$ and satisfies 
  almost everywhere, 
   for any $\psi \in {\cal D}(A)$,
  \begin{eqnarray}
  \int_{{\cal M}} d\mu_{g}(y)\: A^{-1}(x,y) (A \psi)(y) = \psi(x) 
  \:\:\:\:\:\:\: ( \mbox{      namely     } A^{-1} A = I_{{\cal D}(A)})
  \label{g1}
  \end{eqnarray}
   and (almost everywhere),
    for any $\psi \in L^{2}({\cal M}, d\mu_{g})$
  \begin{eqnarray}
   A_{x} \int_{{\cal M}} d\mu_{g}(y)\: A^{-1}(x,y) \psi(y) = \psi(x) 
   \:\:\:\:\:\:\: ( \mbox{      namely     } A A^{-1}  = I) \label{g2}
  \end{eqnarray}
  These relationships are often written in the synthetic  form
  $ A_{x} A^{-1}(x,y) = \delta(x,y)$\\
  In the case $D \geq 4 $, $A^{-1}$ is not Hilbert-Schmidt
  since the 
  series of elements $|\lambda_{j}|^{-2} = \lambda_{j}^{-2}$ diverges
  as follows from {\bf Theorem 1.4}. Anyhow, these 
  facts
  do not forbid the existence of integral kernels,
   which are {\em not} $L^{2}({\cal M}\times{\cal M})$,
  which represent  
  $A^{-1}$ and satisfy either (\ref{g1}) and/or  (\ref{g2}).  
  In any cases, for $P_{0}\equiv 0$, we can state the following  simple
   result. \\
  
  \noindent {\bf Proposition 2.1} {\em Within our general hypotheses
  on  ${\cal M}$ and  $A'$, supposing also $P_{0}\equiv 0$, 
   {\em if } a locally $y$-integrable  $B(x,y)$ exits which satisfy 
  either}  (\ref{g1}) {\em or} (\ref{g2}) {\em with $B(x,y)$ in place of 
  $A^{-1}(x,y)$,  it must be unique (barring differences on vanishing measures
  sets of ${\cal M}\times{\cal M}$)  and the operator $B$ defined by this
  integral kernel must coincide with $A^{-1}$. Furthermore,
  respectively,} (\ref{g2}) {\em or} (\ref{g1}) {\em has to hold true with
   $B(x,y)$ in place of 
  $A^{-1}(x,y)$.}\\
  
  \noindent {\em Proof.}     The uniqueness is a trivial consequence
  of the linearity of (\ref{g1}) or  (\ref{g2}) respectively, 
  taking account that  $R(A)$ is dense in 
  $L^{2}({\cal M},d\mu_{g})$ in the first case, and that A is 
  injective
  in the other case. The coincidence $B=A^{-1}$ follows by 
  pure algebraic considerations straightforwardly. Then, the last point of 
  the thesis
  follows from the definition of inverse operator. $\Box$\\
  
  \noindent {\em Comments.}\\
  {\bf (1)} Dropping the hypotheses of $P_{0}\equiv 0$,
  $A^{-1}$ does not exist and  also
  any integral kernel $B(x,y)$ which satisfies (\ref{g2}) on $R(A)$,
  if exists, cannot be uniquely determined since $B(x,y) + c 
  \phi(x)\phi^{*}(y)$   satisfies the same equation for any $c\in \C$
  whenever  $A\phi = 0 $. (Notice that 
  $\phi$ can be redefined on a set of vanishing 
  measure so that it belongs to $C^{\infty}({\cal M} )$ by {\bf Theorem 
  1.1}).\\ 
  {\bf (2)} We stress that,
   {\bf Proposition 2.1} is much more 
  general, indeed, trivially, it holds true also considering $A$ : ${\cal 
  D}(A)$ 
  $\rightarrow$ ${\cal H}$, where ${\cal D}(A)$ is a dense subspace (not 
  necessarily closed)   of a Hilbert space ${\cal H} = L^{2}(X,d\mu)$,
  $\mu$ being any positive measure on $X$,
  $A=A^{\dagger}$ and $\sigma(A) \subset (0, +\infty)$.\\ 
  {\bf (3)} In particular,  for $Ker$ $A =\{ 0\} $,
  {\bf Proposition 2.1} holds true considering $A^n$,
  for any positive integer $n$, rather than $A$ self, where $A$ is the 
  Friedrichs extension of $A'$ and $A^n$ is defined via spectral 
  theorem as usual.\\
  {\bf (4)} Notice 
that, in this case $A^{-n}$ is trace class if and only if  $n>D/2$
  as consequence of {\bf Theorem 1.4}.\\
  
The following Theorem gives a practical realization of the Green
functions of $A^n$ in terms of $\mu^{-2n}\zeta(n,x,y|A/\mu^2)$.\\

  \noindent {\bf Theorem 2.5.}  {\em Within our  general hypotheses on
  $A'$ and ${\cal M}$, supposing also $r>0$ and $P_{0}\equiv 0$ and 
  fixing any positive integer $n$,}

   (a) {\em $ (x,y) \mapsto
  \mu^{-2n} \zeta(n,x,y| A/\mu^{2})$, 
  not depending on  $\mu^{2}>0$, for 
  $(x,y) \in ({\cal M}\times {\cal M}) - {\cal D}$ 
  (${\cal D} = \{ (x,y) \in  {\cal M} \times {\cal M} \:|\: x=y \}$)
  defines the unique
  Green function of $A^n$, 
  \begin{eqnarray}
  G(x,y| A^{n}) := \zeta(n,x,y|A) \label{g} 
  \end{eqnarray}
   in the sense that this is the unique $C^{0}(({\cal M}\times{\cal 
   M}) -{\cal D})$
   function 
   which satisfies  for any $\psi \in {\cal D}(A^n)$ (${\cal D}(A^n)$ 
   is the domain of $A^n$ obtained from the spectral theorem) 
  \begin{eqnarray}
  \int_{{\cal M}} d\mu_{g}(y)\: G(x,y|A^n) (A^n \psi)(y) = \psi(x) 
 \mbox{      (almost everywhere)   }  \label{g1'}
  \end{eqnarray}
   and, for any $\psi \in L^{2}({\cal M}, d\mu_{g})$
  \begin{eqnarray}
   A^n_{x}
    \int_{{\cal M}} d\mu_{g}(y)\: G(x,y|A^n) \psi(y) = \psi(x)
     \mbox{      (almost everywhere)   }  \label{g2'} 
  \end{eqnarray}
    and thus defines
  the integral kernel of $A^{-n}$.}

  (b) {\em Dropping the hypothesis $P_{0} \equiv 0$, $ (x,y) \mapsto 
  \mu^{-2n} \zeta(n,x,y| A/\mu^{2}) 
  =: G(x,y| A^n) $ 
  does not depend on $\mu^{2}>0$, and} 
  {\em it satisfies} (\ref{g1'})
 {\em anymore for  $\psi \in {\cal D}(A^n)  
 \cap \{ \mbox{Ker} A \}^{\perp}$ and produces
  a vanishing right hand side for $\psi \in$ $Ker$ $A$;}
 {\em and  it satisfies also} (\ref{g2'}) {\em for any $\psi \in R(A^n)$.}

  (c) {\em For $n > D/2$ ($D>1$), in both cases (a) and (b), 
  $G(x,y|A^n)$ does not diverge for $y\rightarrow x$ and 
  is continuous in  ${\cal M} \times {\cal M}$.
  Moreover,
  \begin{eqnarray}
     \int_{{\cal M}} G(x,x|A^{n}) d\mu_{g}(x) = \mbox{Tr} A^{-n}\:,
   \label{traceII}
  \end{eqnarray}
  (where $A^{-n}$ in the the trace is defined via spectral theorem
  dropping the part of the spectral measure on the kernel  of $A$
  whenever it is not trivial).}\\

  \noindent {\em Sketch of Proof.  } (See  also \cite{Ag} for the item (c)) 
  The uniqueness of the Green function in the case (a) is a trivial 
  consequence of {\bf Proposition 2.1} and the remark (3) above. The 
  divergences as $x\rightarrow y$ in the Green functions of $A^n$ given
  in (\ref{g}) can be analyzed employing the truncated local $\zeta$
  function as we done studying the effective Lagrangian. The remaining
  part of the theorem is based on the following identities used recursively,
for any $\psi \in {\cal D}(A)$, where $'$ indicate the $t$
   derivative
    \begin{eqnarray}
   & & \int_{{\cal M}} d\mu_{g}(y)\: \zeta(1, x,y|A) (A\psi)(y)
    = \int_{0}^{+\infty} dt \int d\mu_{g}(y)\: K(t,x,y) (A\psi)(y) \nonumber\\
   &= &  \lim_{\epsilon \rightarrow 0^{+}}
    \int_{\epsilon}^{1/\epsilon}   \int  d\mu_{g}(y)\: K(t,x,y)
    (A\psi)(y) =
    \lim_{\epsilon \rightarrow 0^{+}}  \int_{\epsilon}^{1/\epsilon} dt
     \int d\mu_{g}(y)\: (A_{y} K(t,x,y))
    \psi(y)\nonumber\\
    &=& - \lim_{\epsilon \rightarrow 0^{+}}
    \int_{\epsilon}^{1/\epsilon}dt \int d\mu_{g}(y)\:  K(t,x,y)'
    \psi(y)\nonumber
     = -  \lim_{\epsilon \rightarrow 0^{+}}
     \int_{\epsilon}^{1/\epsilon} dt \left(\int d\mu_{g}(y)\:  K(t,x,y)
    \psi(y)\right)' \nonumber\\
     &=& - \lim_{\epsilon \rightarrow 0^{+}} (e^{-(1/\epsilon) A} \psi)(x)
     + \lim_{\epsilon \rightarrow 0^{+}} (e^{-\epsilon A} \psi)(x)
     = \psi(x)\:. \:\:\:\:\:\:\:\:\:\:\:\:\:\:\:\:\:\:\:\:\:\:\:\:\:\:\:\:\:\:\:\:\:\:\:\:\:\:\:\:\:\:\:\:\: \Box \nonumber
    \end{eqnarray}

   We remark that,
 If $P_{0}$ is untrivial the Green function of $A^{n}$
    is not clearly defined because $A^n$ is not injective, in fact,
     the "Green 
    function" $G(x,y|A^{n})$ defined via local $\zeta$ function 
    (\ref{g}) correspond to a possible  choice for a right-inverse of 
    the operator $A^n$.
     Conversely, 
for $P_{0}\equiv 0$,  $\zeta(n,x,y|A)$ is the unique   Green
    function of $A^n$,  and not 
    $A'^n$. Actually in our case there is no
    ambiguity since $A'$ determines unambiguously its self-adjoint 
    extension $A$. However, all results given within this subsection
     hold true also considering manifolds which are compact with boundary. 
    In that case it is possible to  have different 
     self-adjoint extensions of $A'$ determined by  different 
     boundary conditions one may impose on the functions in ${\cal 
     D}(A')$.
       In such a situation, provided $P_{0}\equiv 0$, the Green
        function is still uniquely determined by the chosen self-adjoint 
       extension of $A'$  or, equivalently, by the chosen boundary 
       conditions.

  \subsection {$\langle \phi^{2}(x|A)\rangle $, point-splitting and
   Hadamard expansion.}

  The procedure of the point-splitting for the field fluctuation 
  is based, once again, upon a
  divergence subtraction procedure in the limit coincidence of the 
  arguments of the Green function
  \begin{eqnarray}
  \langle \phi^{2}(y|A)\rangle_{\mu^{2}} = \lim_{x\rightarrow y}
  \left\{  G'(x,y|A) - \mbox{      "divergences"    } \right\}\:,
   \label{pointphi}
  \end{eqnarray}
  where $G'(x,y|A)$ is a "Green function" of the operator $A$, namely 
  an integral kernel of the operator $A^{-1}$ provided it exists.
 On the physical ground $G'(x,y|A) $
  should determine  the quantum state completely after the 
  "Lorentzian"-time
  analytic continuation \cite{ha,wa,bd,fullingruijsenaars,fu,waldlibro}
  by determining the Feynman propagator as well as the Wightman 
  functions of any order for the quasifree state \cite{kw}. 

  Concerning the "divergences" above,
 we want to  determine them  directly from
  the $\zeta$ function approach assuming (\ref{g}) in the case $n=1$
  as the Green function to put in the expression above. 
  Notice that this identification is automatic whenever $P_{0} \equiv 0$ 
  because of the uniqueness of the Green function proven above.
  Anyhow, we shall assume (\ref{g}), for $n=1$, {\em also}
   in the case $Ker$  $A \neq 
  \{ 0\}$ where the concept of Green function is not so clearly 
  understood. 
  
     Let us proceed as in the case of the effective action. We have, from
   (\ref{dim2''}),
  \begin{eqnarray}
  Z(s,y|A/\mu^{2}) &=& 
  \frac{s}{\mu^{2}}\zeta(s+1,y|A/\mu^{2})\nonumber\\
   &=& \frac{s}{\mu^{2}}\zeta(N,s+1,y,y|A/\mu^{2}, \mu_{0}^{-2})
 - \left(\frac{\mu}{\mu_{0}} \right)^{2s+2}
 \frac{s P_{0}(y,y)}{\mu^{2}(s+1)\Gamma(s+1)}
 \nonumber\\
& & + \frac{s\mu^{2s}}{(4\pi)^{D/2}}
\sum_{j=0}^{N}
\frac{a_{j}(y,y|A) (\mu_{0}^{-2})^{(s+1+j-D/2)}}{\Gamma (s+1) (s+1+j-D/2)}
  \label{aaa}\:.
  \end{eqnarray}
  Using the definition (\ref{phiz}) we have
   \begin{eqnarray}
  \langle \phi^{2}(x|A) \rangle
   &=& \frac{1}{\mu^{2}}\zeta(N,1,y,y | A/\mu^{2}, \mu_{0}^{-2})
 - \frac{P_{0}(y,y)}{\mu_{0}^{2}} 
+\delta_{D}\frac{a_{D/2-1}(y,y)}{(4\pi)^{D/2}}\left[ \gamma + \ln \left(
\frac{\mu}{\mu_{0}}\right)^{2} \right]
 \nonumber\\
& & + \frac{1}{(4\pi)^{D/2}}
\sum_{j=0; j\neq D/2-1}^{N}
\frac{a_{j}(y,y|A) (\mu_{0}^{-2})^{(1+j-D/2)}}{(1+j-D/2)}\:,
\label{middle}
  \end{eqnarray}
 where $\delta_{D} =0$
 if $D$ is odd and $\delta_{D}=1$ otherwise ($D>1$ in both cases). 
 Now, we notice that, from {\bf Lemma 2.1},
  $\zeta(N,1,x,y|A/\mu^{2}, \mu_{0}^{2})$ is 
 continuous for $x\rightarrow y$  since it holds $N>D/2+2$.
We can re-write $\zeta(N,1,x,y|A/\mu^{2}, \mu_{0}^{2})$ in 
the right hand side of (\ref{middle}) by employing (\ref{dim''})
 \begin{eqnarray}
\zeta(N,1,y,y| A/\mu^{2}) &=& 
 \lim_{x\rightarrow y} \left\{ \zeta(1,x,y|A/\mu^{2}, \mu_{0}^{-2})
 + \left(\frac{\mu}{\mu_{0}} \right)^{2} P_{0}(x,y) \right.
 \nonumber\\
& &\left.  -\frac{\mu^{2}}{(4\pi)^{D/2}}
 \sum_{j=0}^N a_j(x,y|A)\int_{0}^{\mu_{0}^{-2}} dt\: t^{j-D/2} 
{e^{-\sigma(x,y)/2t}} \right\}  \:.\nonumber
\end{eqnarray}  
 Inserting this result in (\ref{middle}), we get
 \begin{eqnarray}
 \langle \phi^{2}(y|A) \rangle 
&=& \delta_{D}\frac{a_{D/2-1}(y,y|A)}{(4\pi)^{D/2}}
 \left[ \gamma + \ln \left(
\frac{\mu}{\mu_{0}}\right)^{2} \right]\nonumber \\
&+&  \lim_{x\rightarrow y} \left\{ 
\zeta(1,x,y|A) +
 \sum_{j=0; j\neq D/2-1}^{N}
  \frac{a_{j}(x,y|A)}{(4\pi)^{D/2}}
\left[ \frac{\mu_{0}^{-2(j-D/2+1)}}{j-D/2+1} \right. \right. \nonumber \\
&-& \left.\left(\frac{\sigma}{2}\right)^{j-D/2+1}
\int_{\sigma\mu^{2}_{0}/2}^{+\infty}  du\: u^{D/2-j-2} 
e^{-u}\right] \nonumber\\
&-& \left. \delta_{D} \frac{a_{D/2-1}(x,y|A)}{(4\pi)^{D/2}}  
\int_{\sigma \mu^{2}_{0}/2}^{+\infty}
  du\: \frac{e^{-u}}{u} \right\}  \label{finephi}\:.
  \end{eqnarray}
Above, $N$ is a fixed integer and 
 $N \geq D/2+2$. The integrals above have been still  
computed in (\ref{pri}) -- (\ref{ul}) since
 \begin{eqnarray} 
\frac{a_{j+1}(x,y|A)}{2(4\pi)^{D/2}}
 \left(\frac{\sigma}{2}\right)^{j-D/2+1}
 \int_{\sigma\mu^{2}_{0}/2}^{+\infty}  du\: u^{D/2-j-2} 
e^{-u} = 
H_{j+1}(\sigma)\:.
 \end{eqnarray}
Using  (\ref{pri}) -- (\ref{ul}) in (\ref{finephi}), we finally get
the following theorem.\\

\noindent {\bf Theorem 2.6.} {\em Within our general hypotheses on ${\cal M}$
and $A'$, for $r>0$ and $D>1$ the field fluctuation computed by} 
(\ref{phiz})
{\em can be computed also by a point-splitting procedure for $\langle 
\phi^{2}(y|A) \rangle $. Indeed,
\begin{eqnarray}
\langle \phi^{2}(y|A) \rangle_{\mu^{2}} &=&  
\frac{2\gamma a_{D/2-1}(y,y|A)  }{(4\pi)^{D/2}} \nonumber\\ 
& &+
\lim_{x\rightarrow y} \left\{ G(x,y|A)   -\sum_{j=0}^{D/2-2} (D/2-j-2)! 
\left(\frac{2}{\sigma}\right)^{D/2-j-1} \frac{a_{j}(x,y|A)}{(4\pi)^{D/2}} 
 \right. 
\nonumber\\ 
& & \left.
+ \frac{a_{D/2-1}(x,y|A)}{(4\pi)^{D/2}} \ln\left( \frac{\sigma \mu^{2}}{2}
\right)
\right\}  \label{phipari}\:, 
\end{eqnarray}   
whenever $D$ is even.  The term containing the sum over $j$ appears for 
$D\geq 4$ only.
\begin{eqnarray}
\langle \phi^{2}(y|A) \rangle_{\mu^{2}} &=&  
\lim_{x\rightarrow y} \left\{ G(x,y|A) 
 -\sum_{j=0}^{(D-5)/2} 
\frac{(D-2j-4)!! \sqrt{\pi}}{2^{(D-3)/2-j}} 
\left(\frac{2}{\sigma}\right)^{D/2-j-1} 
\frac{a_{j}(x,y|A)}{(4\pi)^{D/2}} \right.\nonumber \\
 & &-\left.  \frac{a_{(D-3)/2}(x,y|A)}{(4\pi)^{D/2}} 
\sqrt{\frac{2\pi}{\sigma}} \right\}  \label{phidispari}\:, 
\end{eqnarray} 
whenever $D$ is odd. 
The term containing the sum over $j$ appears for $D\geq 5$
only}.  \\

\noindent {\em Comments.}\\
{\bf (1)} First of all, notice that $\mu^{2}$ has disappeared from the final 
result in the case $D$ is odd. Once again, the only task of $\mu^{2}$
is to make physically sensible the argument of the logarithm in the 
case $D$ is even.\\
{\bf (2)} Eq. (\ref{phipari}) and (\ref{phidispari}) prove
that $G(x,y|A)$ has the Hadamard singular behaviour \cite{garabedian}
for $x \sim y$. Indeed, the terms after $G(x,y|A)$ in the right 
hand sides of the equations above, taking account that 
$\langle \phi^{2}(y|A) \rangle = Z'(0,y|A)$ is a regular function 
of $x$ due to {\bf Theorem 2.4}, give the singular part of $G(x,y|A)$
which is just that considered in building up perturbative Hadamard's local 
fundamental solutions \cite{garabedian}.
On the physical ground,
 this means that the quantum state associated to the Green function
 is Hadamard at least in the Euclidean section of the manifold.
 This is a very important point concerning the stress tensor 
 renormalization procedure \cite{kw,waldlibro} by the point-splitting 
 approach.
 Generally speaking, the {\em point splitting procedure} as known 
from the literature (see 
 \cite{bd,wa,fu,waldlibro}
 and references therein) consists of subtracting, from
the Green function, a Hadamard local solution, namely, a $C^{\infty}$ function
of $x$ defined  in a normal convex neighborhood of $y$ of the form 
\begin{eqnarray} 
H_{LMN}(x,y) = \Theta_{D}\frac{U_{L}(x,y)}{ (4\pi)^{D/2}
(\sigma/2)^{D/2-1}}
+ \delta_{D} V_{M}(x,y) \ln (\sigma/2) +  \delta_D 
W_{N}(x,y) \label{Hadamard}
\end{eqnarray}
Where 
$U_{L}(x,y) = \sum_{j=0}^{L} u_{j}(x,y) \sigma^{j}$,
$V_{M}(x,y) = \sum_{j=0}^{M} v_{j}(x,y) \sigma^{j}$ and
$W_{N}(x,y) = \sum_{j=0}^{N} w_{j}(x,y) \sigma^{j}$,
$\delta_{D}$ was defined previously. Moreover,
 $\Theta_{D} :=1$ for  $D \neq 2$ and $\Theta_{2}:=0$. 
All sums above are truncated to some orders $L,M,N $, in particular 
$L = D/2-2$ when $D$ is even.  The corresponding series: 
$M,N \rightarrow \infty $ for $D$ even 
and in $L\rightarrow \infty $ for $D$ odd, generally
 diverges. Actually, concerning our procedure,  it is sufficient taking 
account of the divergent and finite terms 
for $x\rightarrow y$ in the formal series above.\\
There are recursive differential equations, obtained by considering 
the formal 
equation $A_x H=0$ where $H = H_{\infty,\infty,\infty}$,
 that determine  each term of the expansions above. In particular,
the coefficients   of the expansions of
$U$ and $V$ are completely determined  by requiring that
 $H_{\infty,\infty\infty}$, formally, is a Green function of $A$. 
This means that one has to fix opportunely the value of the coefficient 
of the leading divergence as $x \rightarrow y$ as said in the end of Section 2
in  Chapter 5 of \cite{garabedian}. In the practice, with our definition of 
the Riemannian measure, it must be 
$u_0(y,y) = 4 \pi^{D/2}/ [D(D-2) \omega_D]$ where $\omega_D$ is the 
volume of the unitary $D$-dimensional disk
for $D \geq 3$, and $v_0(y,y) = 1/(4\pi) $ for $D=2$. 
Similarly, the coefficients of the formal series for $W_N$ are determined,  
for $j>0$ only, once $w_{0}$ has been fixed. $w_{0}(x,y)$ self can be fixed 
arbitrarily. \\ We have proven by {\bf Theorem 2.6} that the local $\zeta$  
function procedure makes the same  job  made by the point splitting procedure.
In particular, by a direct comparison between equations (\ref{eq1})
(\ref{eq2}) and the equations for $u_{i}, v_{j}, w_{j}$ given in 
Chapter 5 Section 2 of 
\cite{garabedian}, and by a comparison between 
$u_0(x,y)$ and the corresponding
terms in (\ref{phipari}) and (\ref{phidispari}), 
 one can check straightforwardly that the procedure
pointed out in 
(\ref{phipari}) and (\ref{phidispari}) consist of the coincidence
limit of the difference between the Green function and Hadamard 
solutions with, respectively, $L=D/2-2$, $M=0$, $N=0$ for any even  $D>1$
and $L=D/2-3/2$ for any odd  $D>1$.   
Moreover, whenever $D$ is even, the Hadamard solution is completely determined
 by choosing 
\begin{eqnarray}
w_{0}(x,y) = -\frac{a_{D/2-1}(x,y|A)}{(4\pi)^{D/2}} (2\gamma +\ln \mu^{2})
\end{eqnarray}
Therefore, the local $\zeta$ function procedure  picks out particular 
Hadamard solutions by a particular choice of 
$w_{0}(x,y)$. We stress that, within the point splitting 
procedure for the field fluctuations, it seems that there is no general way 
to choose a particular function $w_{0}(x,y)$ rather than another one
(the situation is a bit different concerning the {\em stress tensor} where 
one can impose other constraints as the conservation of the renormalized 
stress tensor). 
Obviously there is no guarantee that the
choice of $w_{0}(x,y)$ performed by the  $\zeta$ function
procedure is the physical one (if it exists).\\
Finally, we remark that the obtained result should hold also in the case
${\cal M}$ is not compact. In this case, in general, $A'$ admits different
self-adjoint extensions and thus Green functions,
corresponding to different physical states. The important point is that
the Hadamard expansion does not depend on the considered  self-adjoint 
extension. In other words, the singularity eliminated
by the point-splitting procedure from the Green function 
is universal, depending on the local geometry only.\\
{\bf (3)} In the general case, similarly to the results found for the
      effective Lagrangian, an ambiguity appears because
the presence of the scale $\mu$ (for $D$ even). In fact, any rescaling
$\mu^2 \rightarrow \alpha \mu^2 $ changes the value of $\langle \phi^2(x|A)
\rangle_{\mu^2}$ producing 
\begin{eqnarray}
\langle \phi^2(x|A) \rangle_{\alpha\mu^{2}}
 = \langle \phi^2(x|A) \rangle_{\mu^{2}} + 
\delta_{D}\frac{a_{D/2-1}(x,x|A)}{(4\pi)^{D/2}}   \ln \alpha \:. 
\end{eqnarray}
 It is worth stressing that this ambiguity concerns just 
 the term $w_{0}(x,y)$ of the Hadamard expansion (see point (2) above).
 Therefore, in a general approach containing both local $\zeta$ function
  regularization and point-splitting procedure, the ambiguity pointed out above
 can be generalized into 
\begin{eqnarray}
\langle \phi^2(x|A) \rangle_{\alpha\mu^{2},\delta w_{0}}
 = \langle \phi^2(x|A) \rangle_{\mu^{2}} + 
\delta_{D}\frac{a_{D/2-1}(x,x|A)}{(4\pi)^{D/2}}  
 \ln \alpha - \delta_D \delta w_{0}(x,x) \:,
\end{eqnarray}
$\delta w_{0}(x,y)$ being any smooth function  in ${\cal M}\times {\cal M}$ 
which could  be fixed by imposing some 
further physical constraints.

  \subsection{Further properties of $Z(s,x|A/\mu^{2})$ and $\langle 
  \phi^2(x|A)\rangle $.}

  Within this section we want to prove a local version of a 
  formula  related to the field fluctuation and to the change of the 
  mass in the field operator. 
  Concerning the $\zeta$ function, 
  similar formulae have appeared in \cite{ca} with the hypothesis that  
  ${\cal M}$ is a homogeneous  space, and in \cite{rep} for the  integrated
  $\zeta$ function without a rigorous proof.
  Concerning the field fluctuations, similar formulae for the 
  particular case of  homogeneous four-dimensional spaces can be found 
  in \cite{dm}.  Here, we shall deal with much more general hypotheses. \\

  \noindent {\bf Theorem 2.7} {\em Within our general hypotheses on ${\cal 
  M}$ and $A'$, and supposing 
  $r>0$,  let $\lambda$ be the first nonvanishing eigenvalue of $A$.
  For any real $\delta m^{2}$ such that  $0< \delta m^{2} <\lambda$ 
  and any integer $K>0$,  posing $B' := A' + \delta m^{2} I$, one has
  \begin{eqnarray} 
  \zeta(s,x|B/\mu^{2}) &=& \left( \frac{\mu^{2}}{\delta 
  m^{2}}\right)^s P_{0}(x,x) +
  \sum_{n=0}^{K} \left(-\frac{\delta m^{2}}{\mu^{2}}\right)^n 
  \frac{\Gamma(s+n)}{n!\Gamma (s)}\zeta(s+n,x|A/\mu^{2}) \nonumber \\ 
  & & + \sum_{n=K+1}^{+\infty} \left(-\frac{\delta m^{2}}{\mu^{2}}\right)^n 
  \frac{\Gamma(s+n)}{n!\Gamma (s)}\zeta(s+n,x|A/\mu^{2}) \:, 
  \label{fine}\
  \end{eqnarray}
   where  $x\in {\cal M}$ is fixed and 
  $Re$ $s\in [D/2-K,  +\infty)$. Furthermore, the convergence of the 
  series is   uniform in any set
  $Re$ $s\in [D/2-K,  \beta]$ for 
  any real $\beta >D/2-K$.}\\
  
  \noindent {\em Proof.} See {\bf Appendix}. $\Box$\\
  
  There is a trivial corollary of the theorem above concerning 
  the field fluctuations. Indeed, we have for $Z(s,x|A/\mu^{2})$ by 
  (\ref{phiz'})
  \begin{eqnarray} 
  Z(s,x| A+ \delta m^{2}I)/\mu^{2}) &=& \frac{s}{\delta m^2}
  \left( \frac{\mu^{2}}{\delta 
  m^{2}}\right)^{s} P_{0}(x,x) + Z(s,x | A/\mu^{2})\nonumber \\
  &+&  \sum_{n=1}^{K} \left(-\frac{\delta m^{2}}{\mu^{2}}\right)^n 
  \frac{s\Gamma(s+1+n)}{\mu^{2}
  \Gamma(n+1)\Gamma (s+1)}\zeta(s+n+1,x|A/\mu^{2}) \nonumber \\ 
  &+& \sum_{n=K+1}^{+\infty} \left(-\frac{\delta m^{2}}{\mu^{2}}\right)^n 
  \frac{s\Gamma(s+n+1)}{\mu^{2}\Gamma(n+1)\Gamma (s+1)}
  \zeta(s+n+1,x|A/\mu^{2}) \:. \nonumber
   \end{eqnarray}
  Notice that we can take the $s$ derivative of 
this identity for $s=0$ passing the 
  derivative under the symbol of series because each term of the series 
  above is analytic and the series converges uniformly
  provided $K > D/2 -1$ (the $-1$ is due to the evaluation
  of $\zeta(s,x)$ for $s+1$ in order to get $Z(s,x)$). \\
     By (\ref{phiz}) we have
  \begin{eqnarray} 
  \langle \phi^{2}(x|A+ \delta m^{2}I)\rangle_{\mu^{2}} 
  &=& \langle \phi^{2}(x|A)\rangle_{\mu^{2}} + 
   \frac{P_{0}(x,x)}{\delta m^2} \nonumber \\
  &+& \sum_{n=1}^{+\infty} \left(-\frac{\delta m^{2}}{\mu^{2}}\right)^n 
  \left[\frac{s\Gamma(s+1+n)}{\mu^{2}
  \Gamma(n+1)\Gamma (s+1)}\zeta(s+n+1,x|A/\mu^{2})\right]'_{s=0} \nonumber  
  \end{eqnarray}
  where the prime means the $s$ derivative.
  By the point (b) of {\bf Theorem 2.2} and 
  the point (c) of {\bf Theorem 2.5} we can rewrite the formula 
  above in a improved form.\\
  
\noindent {\bf Theorem 2.8.}  {\em In the same hypotheses of }
{\bf Theorem 2.7} {\em the field fluctuations evaluated via local $\zeta$
  function approach for the operator $A$ and $A+ \delta m^{2} I$
  are related by the relation ($n$ is integer)} 
  \begin{eqnarray} 
  \langle \phi^{2}(x|A+ \delta m^{2}I)\rangle_{\mu^{2}} 
  &=& \langle \phi^{2}(x|A)\rangle_{\mu^{2}} + \sum_{1\leq n \leq D/2-1} 
   (-\delta m^2)^n \Phi_{n}(x|A)_{\mu^{2}} \nonumber \\ 
  & + &  \sum_{n > D/2-1}
   (-\delta m^{2})^n   G(x,x|A^{n+1})    \frac{P_{0}(x,x)}{\delta m^{2}} 
\:.\label{fine'''}
  \end{eqnarray}
   {\em where, if  $D$ is odd $\Phi_{n}(x|A)_{\mu^{2}}$ does not 
   depend on $\mu^{2}$ and} 
  \begin{eqnarray}
 \Phi_{n}(x|A)_{\mu^{2}} = \mu^{-2(n+1)} \zeta(n+1,x|A/\mu^{2}) =
 \zeta(n+1,x|A)
  \end{eqnarray}
 {\em and, if $D$ is even}
  \begin{eqnarray}
 \Phi_{n}(x|A)_{\mu^{2}} = \mu^{-2(n+1)} 
 \left[\frac{s\Gamma(s+1+n)}{\Gamma(n+1)\Gamma (s+1)}
 \zeta(s+n+1,x|A/\mu^{2})\right]'_{s=0} 
 \end{eqnarray}
  
   Notice that $\Phi_{n}(x|A)_{\mu^{2}}$ is always well-defined due to 
   the meromorphic structure of the local zeta function
   which involves simple poles only. Moreover,
    (\ref{fine'''}) holds true also in the 
  case $P_{0}$ does not vanish and thus $A^{-1}$ does not exit. In 
  this case
  the "Green functions" $G(x,y|A^{n+1})$ are not uniquely determined 
  and are  those defined via $\zeta$ by (\ref{g}).

  The found relation is the mathematically correct {\em local} form, in 
  closed manifolds, 
    of a formal relation assumed by physicists \cite{ev}, namely,
     \begin{eqnarray}
   \int_{{\cal M}} \langle \phi^{2}(x|A+ \delta m^{2}I)\rangle_{\mu^{2}} 
   d\mu_{g}(x)
  &=& \int_{{\cal M}} \langle \phi^{2}(x|A)\rangle_{\mu^{2}}  d\mu_{g}(x)
  + \sum_{n=1}^{+\infty} (-\delta m^{2})^n \mbox{Tr} A^{-(n+1)} \:. 
  \label{incorrect}
  \end{eqnarray}
  To get 
  (\ref{incorrect})  one starts from the {\em correct}
  expansion holding for $|\delta m^{2}| < || A^{-1} ||^{-1}$ (provided 
  $A^{-1}$ exists)
  \begin{eqnarray}
  (A + \delta m^{2} I)^{-1} = A^{-1} + \sum_{n=1}^{+\infty} (-\delta 
  m^2)^{n} A^{-(n+1)}
  \end{eqnarray}
  and uses  the linearity of the trace operation and 
  the generally {\em incorrect} identities  ($n=1,2,\ldots $)
  \begin{eqnarray}
  \langle \phi^{2}(x|A)\rangle  =  G(x,x|A) \:\:\:\:\:\: \mbox{and}
\:\:\:\:\:\:
  \mbox{Tr} A^{-n} = \int_{{\cal M}} G(x,x|A^n) 
   d\mu_{g}(x)
  \nonumber
  \end{eqnarray}
   The identities above do not hold in every cases as pointed out 
   previously.  In particular, barring trivial cases,
   the physically relevant dimension $D=4$ generally involves
   the failure of both the identities above.   
   Actually, the former identity never holds for $D>1$.
   The latter generally
   does not hold for $n\leq D/2$ because $A^{-n}$ is not a trace class
   operator.  For $D=4$, problems arise for the term $n=1$
    in (\ref{incorrect}). The task of the first sum in the right hand
   side of (\ref{fine'''}) is just to regularize the failure of the
   second identity above.     In the case $D=4$, (\ref{fine'''}) reads
   \begin{eqnarray} 
  \langle \phi^{2}(x|A+ \delta m^{2}I)\rangle_{\mu^{2}} 
  &=& \langle \phi^{2}(x|A)\rangle_{\mu^{2}} + 
   \frac{P_{0}(x,x)}{\delta m^{2}} + 
  (-\delta m^2)\Phi_{1}(x|A)_{\mu^{2}} \nonumber \\ 
  & + &  \sum_{n > 1}
   (-\delta m^{2})^n   G(x,x|A^{n+1}) \:.\label{fineI'''}
  \end{eqnarray}
   We have also, after trivial calculations
     \begin{eqnarray}
 \Phi_{1}(x|A)_{\mu^{2}} = 
 \left[ s\zeta(s+2,x|A)\right]_{s=0}
 +\mu^{-4}  \left[ s \zeta(s+2,x|A/\mu^{2})\right]'_{s=0}\:.  
 \end{eqnarray}
 A final remark for $D=4$ is that, as one can prove directly 
  \begin{eqnarray}
  \int_{{\cal M}}  \left[ s\zeta(s+2,x|A)\right]_{s=0} d\mu_{g}(x)
  = \left[ \int_{{\cal M}}  
   s\zeta(s+2,x|A)
   d\mu_{g}(x)\right]_{s=0} = \left[ s\zeta(s+2|A)\right]_{s=0} \label{mc}
   \end{eqnarray}
   In the case $D=4$ and $P_{0}\equiv 0$, 
   one can check that $A^{-2}$ is Hilbert-Schmidt
   and thus compact, moreover it is not trace-class but it belongs to 
      ${\cal L}^{1+}$, 
   the Macaev ideal \cite{connes}.   
   Let us further suppose that $A$ is a pure Laplacian.
   Then,  the last term in the right hand side of (\ref{mc}) 
   is nothing but  the Wodzicki
   residue of $A^{-2}$ \cite{wo,connes}. In other words, in the 
   considered case,
   the last term in the right hand side of (\ref{mc}) is four times 
   the Dixmier trace of $A^{-2}$ \cite{dix}
    because of a known theorem by Connes
   \cite{connes,elw}. 

   We conclude this section noticing that,
 differently to that argued in \cite{ev},
    not only the local $\zeta$ function approach is consistent, 
   but it also agrees with the point-splitting procedure and 
    it is able to regularize and give a mathematically
   sensible meaning to formal identities handled by 
   physicists\footnote{This reply  concerns only a part of  criticism
   developed in \cite{ev}. Several other papers (e.g. see \cite{reply})
   have recently  appeared to reply to the objections aganist the 
   multiplicative anomaly.}.

 \section{Summary and outlooks.}  
  
  In this paper, we have proven 
  that the local $\zeta$ function technique is rigorously founded  
and produces  essentially the same results of the point-splitting,
 at least considering the 
  effective Lagrangian and the field fluctuations.
  This result holds for any dimension $D>1$ and in closed manifolds
  for Friedrichs extensions $A$ of  Schr\"{o}dinger-like real 
  positive smooth operator $A'$.
  Since these results are {\em local} results, we expect that this 
  agreement does hold  also dropping the hypothesis of a compact 
  without boundary manifold. Several comments toward this generalization 
  have been given throughout the paper.
  
  Differences between the two approaches 
  arise in the case of a untrivial $Ker A$, 
  when the local $\zeta$ function approach can be successfully employed 
  whereas the point-splitting procedure is not completely well-defined.
  
  Another results obtained in this paper 
  is that the two-point functions, namely the  Green
  function  of $A$ which we have built up  via local $\zeta$ function 
  and which is unique 
  provided $P_{0}\equiv 0$, has
  the Hadamard behaviour for short distance of the arguments  for 
  any $D>1$. This fact allows the substantial 
  equivalence of the two methods concerning  the field fluctuation 
  regularization.
   The only difference between the two approaches consists of the 
  different freedom/ambiguity  in choosing the term $w_{0}(x,y)$ 
  of the Hadamard local solution. 
  
  Finally, we have discussed and rigorously proven a particular
  formula concerning the field fluctuations within our approach,
  proving that the $\zeta$ function procedure is able to regularize
  an identity which is supposed true by physicists but involves
  some mathematical problems when one tries to give  rigorous 
  interpretations of it .
 
  An important issue which remains to be investigated is the 
 equivalence of the 
  local $\zeta$-function approach and the point-splitting one
   concerning the one-loop stress tensor. This is an intriguing
   question also because the   following weird reason.
    The point-splitting 
  approach does not work completely in its naive formulation, as pointed out 
  in \cite{waldlibro} (see also \cite{bd,fu}), at least 
  in the case of a massless  scalar
  field. In this case one cannot use Schwinger-DeWitt algorithm to
  pick out the term $w_0$ in the Hadamard expansion and, putting 
 $w_0 \equiv 0$ one  has to adjust by hand the final result 
  to get either the conservation of the obtained stress tensor and,
in the case of conformal coupling, the 
  appearance of the conformal anomaly.
  Actually, this drawback does not arise within the local $\zeta$
  function approach, as pointed out in \cite{moa}, because the method
 does not distinguish between different values of the mass and the coupling
 with the curvature.\\

\noindent{\em Acknowledgment.}
I am  grateful to A. Cassa, E. Elizalde, E. Pagani, L. Tubaro 
 and S. Zerbini  for useful discussions and suggestions  about several topic 
 contained in this paper. 
I would like to thank   R. M. Wald who pointed out \cite{wa} to me. 
This work has been financially supported by a Research Fellowship 
of the Department of Mathematics of the Trento University.

\section*{Appendix: Proof of some theorems.}

\noindent {\bf Proof of Theorem 2.2.}
 The idea is to break off the 
integration in (\ref{zetad}) for $Re$ $s>D/2$ as
\begin{eqnarray}
\zeta(s,x,y|A/\mu^2) &=& \frac{\mu^{2s}}{\Gamma(s)}
\int_0^{+\infty} dt\: t^{s-1} 
\left[ K(t,x,y|A) - P_0(x,y|A) \right] \\
&=&  \frac{\mu^{2s}}{\Gamma(s)} \int_{0}^{\mu_{0}^{-2}} 
\left\{\ldots\right\} + \frac{\mu^{2s}}{\Gamma(s)} 
\int_{\mu_{0}^{-2}}^{+\infty} 
\left\{  \ldots  \right\}\:,  \label{breack}
\end{eqnarray}
where  $\mu_{0}>0$ is an arbitrary mass cutoff.\\
 We  study the properties of these integrals separately.
Let first focus attention on the second integral in right hand side of 
(\ref{breack}) considered as a function of $s\in \C$, $x,y \in {\cal M}$.\\
From {\bf Theorem 1.1}, and using Cauchy-Schwarz inequality, one finds 
straightforwardly  
\begin{eqnarray}
|K(t,x,y|A) - P_0(x,y|A)|^{2}
  \leq [K(t,x,x|A) - P_0(x,x|A)] [K(t,y,y|A) - P_0(y,y|A)]\:.
\end{eqnarray}
Moreover, let us define 
\begin{eqnarray}
 p(t,x) := e^{\lambda t } [K(t,x,x|A) - P_0(x,x|A)] \geq 0 \label{brutta}
 \end{eqnarray}
$\lambda $ being the first strictly positive 
eigenvalue of $A$. From the expansion (\ref{sum 0})
it is obvious that, since $\lambda_{j} - \lambda \geq 0 $ whenever 
$\lambda_{j} \neq 0 $,
 $p(t,x) \geq p(t',x)$ for $t'\geq t$ and thus, in $[\mu_0^{-2}, 
 +\infty)$, 
$p(t,x) \leq p(\mu^{-2}_{0},x)$. This last function is
 continuous in ${\cal M}$ by construction (essentially, it is 
 $K(t,x,x)$ self).  
Hence, $\max_{x\in{\cal M}} p(\mu^{-2}_{0},x)$
does exist because the compactness of ${\cal M}$.
 From 
(\ref{brutta}) one has the $(x,y)$-uniform 
upper bound of the heat kernel in 
$t\in [\mu_0^{-2}, +\infty)$
 \begin{eqnarray}
|K(t,x,y|A) - P_0(x,y|A)| 
\leq  \max_{z\in{\cal M}}[K(\mu_{0}^{-2},z,z|A) - P_0(z,z|A)]\:\:  
e^{-\lambda (t- \mu_0^{-2})} 
\label{magg} 
\end{eqnarray}
 This result proves that the second integral in (\ref{breack}) converges 
absolutely, 
not depending on $s\in \C$, and the final function is in $C^0({\cal 
M}\times {\cal M})$ for any fixed $s\in \C$ because the upper bound in 
(\ref{magg}) does not depend on $x,y$.
Actually, studying the function $t \mapsto t^s \exp{(-\lambda t)}$, one 
finds that, for $\mu_{0}>0, \lambda>0, \epsilon \in (0,1)$
 there is a positive constant 
$B=B(\mu_{0}^{-2},\lambda,\epsilon)$ such 
that,  for $s\in \C$, 
$|t^s \exp{(-\lambda  t)}| \leq \exp{[B\mbox{ }  (Re \mbox{ } s)^{2} ]}
\exp{(-\epsilon \lambda t)}$ whenever $t \in [\mu^2_{0}, +\infty)$ and thus 
one has, for $t$ belonging to that interval and $Re$ $s \in [\alpha, 
\beta]$, $\alpha, \beta \in \R $, $\alpha \leq \beta $     
\begin{eqnarray}
| t^s e^{(-\lambda t)} | \leq \left(  e^{B \alpha^{2}} + e^{B 
\beta^{2} } \right) e^{-\epsilon\lambda} \label{magg2}\:. 
\end{eqnarray}
This implies that the considered integral defines also a function of 
$s,x,y$ which belongs to  $C^0( \C \times  {\cal M}\times {\cal M})$.
The same bound (the $s$ derivatives produces integrable factors 
$\ln t$ in the integrand)  proves that the considered integral is also
an analytic function of $s\in \C$ for $x,y$ fixed in ${\cal M}$.
This is because, by Lebesgue's dominate 
convergence theorem, one  can  interchange the symbol 
of $t$ integration with the derivative in $Re$ $s$
 and $Im$ $s$ and  prove Cauchy-Riemann's identities. 
 Moreover, by the same way, it is trivially proven that  all of $s$ 
 derivatives 
 of the considered integral belong to $C^0(\C \times {\cal M}\times {\cal 
 M})$.\\
 The remaining integral in (\ref{breack}) needs further manipulations. 
Using {\bf Theorem 1.3} and 
 (\ref{expansion1}) for any integer  $N> D/2 +2$, it is now convenient to consider
the function
\begin{eqnarray}
\zeta(N,s,x,y|A/\mu^{2}, \mu_{0}^{-2}) &:=& 
 \frac{\mu^{2s}}{\Gamma(s)}
\int_{\mu_{0}^{-2}}^{+\infty} dt\: t^{s-1} 
\left[ K(t,x,y|A) - P_0(x,y|A) \right]  \nonumber \\
& & + \frac{\mu^{2s}}{\Gamma(s)}
 \int_{0}^{\mu_{0}^{-2}} dt\: t^{s-1} 
 \frac{e^{-\eta\sigma(x,y)/2t}}{(4\pi t)^{D/2}}
t^{N} O_\eta(t;x,y)   \label{s}\:. 
\end{eqnarray}
Let us study this function for a fixed $N > D/2+2$. 
For $Re$ $s\in [D/2-N+\epsilon , \beta]$, $\beta$ being any real $> D/2 
-{N} +\epsilon$ and $\epsilon > 0$ another real,
because  $O_\eta(t,x,y)$ is bounded 
 we have the following $(s,x,y)$-uniform 
 bound of the integrand in the second 
 integral in (\ref{s})
\begin{eqnarray}
| e^{-\eta\sigma(x,y)/2t} t^{s-1+N-D/2} O_\eta(t;x,y) |  \leq K t^{\epsilon -1}
\chi_{1}(t) + K \mu_{0}^{-2|\beta +N-D/2-1|} \chi_{2}(t)\:, 
\label{k'servi}
\end{eqnarray}
where $K>0$ is a constant not depending on $x$ and $y$, $\chi_{1}$ is the 
characteristic function of the set $[0,\min \{1, \mu_{0}^{-2}\}]$ 
and $\chi_{2}$ the characteristic function of the set 
$[\min \{1, \mu_{0}^{-2}\}, \mu_{0}^{-2}]$ ($\chi_{2} \equiv 0$ if this
 set is empty).
As before, this result implies that the second integral in (\ref{s}) 
and all of its $s$ derivatives,
considered as a function of $s,x,y$, belong to  
$C^0(\{ s\in \C \mbox{ } | \mbox{ } Re \mbox{ } s > D/2 -N \}\times {\cal M}
 \times {\cal M})$. Moreover, for $x,y$ fixed, this function is 
analytic in $\{ s\in \C \mbox{ } | \mbox{ } Re \mbox{ } s > D/2 -N \}$.\\
Notice that $\zeta(N,0,x,y|A/\mu^{2}, \mu_{0}^{-2}) = 0$ for any
integer $N>D/2 +2$ and $x,y\in{\cal M}$.\\
If the points $x$ and $y$ do not coincide the exponential function 
in the right hand side of  (\ref{s}) sharply decays as $t\rightarrow 
0^+$. In fact, for $(x,y)\in {\cal G}$ where ${\cal G}$ is any compact 
subset of ${\cal M}\times {\cal M}$ 
which does not contain elements of the form 
$(x,x)$, we have the following $(s,x,y)$-uniform 
bound for $Re$ $s\in [\alpha, \beta]$ for any 
choice of $\alpha<\beta$ in $\R$   
\begin{eqnarray}
| e^{-\eta\sigma(x,y)/2t} t^{s-1+N-D/2} O_\eta(t;x,y) | 
& \leq & e^{-\eta\sigma_{0}/2t} 
K' t^{\alpha +N -D/2 -1}
\chi_{1}(t) \nonumber\\ 
& & + K' e^{-\eta\sigma_{0}/2t} \mu_{0}^{-2|\beta +N-D/2-1|} 
\chi_{2}(t)\:,\label{k''servi}
\end{eqnarray}
where $K'>0$ is a constant not depending on $x$ and $y$ and 
$\sigma_{0} = \min_{{\cal G}} \sigma(x,y)$ which is 
strictly positive.
Notice that no limitations appears on the choice of $[\alpha, \beta]$
namely, in the range of $s$.\\
Therefore, by the same way followed in the general case 
we have that $\zeta(N,s,x,y|A/\mu^{2}, \mu_{0}^{-2}) $
defines a function which belongs, together with all of $s$ derivatives,
to $C^{0}(\C\times (({\cal M}\times {\cal M})-{\cal D}) )$ where 
${\cal D}:= \{ (x,y)\in {\cal M}\times{\cal M}\mbox{ }|\mbox{ } x = 
y\} $.\\
 Coming back to the local $\zeta$ function,
by the given 
definition  we have that, for $N>D/2+2$ 
\begin{eqnarray}
\zeta(s,x,y| A/\mu^{2}) &=&  \zeta(N,s,x,y|A/\mu^{2}, \mu_{0}^{-2})
 - \frac{\mu^{2s}P_{0}(x,y)}{\Gamma(s)}\int_{0}^{\mu^{-2}_{0}} dt 
 \:t^{s-1}
 \nonumber\\
& & + \frac{\mu^{2s}}{\Gamma(s)} \int_{0}^{\mu_{0}^{-2}} dt\: t^{s-1} 
\frac{e^{-\sigma(x,y)/2t}}{(4\pi t)^{D/2}}
\chi (\sigma(x,y)) \sum_{j=0}^N a_j(x,y|A) t^j \label{dim} 
 \end{eqnarray}
Up to now  we have proven that, for $x,y$ fixed,
the first term in the right hand side can be analytically continued
(actually is directly computable there) at least in  the set 
$\{ s\in \C \mbox{ } | \mbox{ } Re \mbox{} s > D/2 -N \}$ and 
furthermore  everywhere 
for $x\neq y$, moreover,  varying also $x$ and $y$ one gets a (at least)
$(s,x,y)$-continuous function 
also considering the $s$ derivatives. This function vanishes for 
$s=0$.\\
The 
second term in the right hand side of (\ref{dim}) can be computed for $Re$ $s>0$ and then the result can be 
continued in the whole $s$ complex plane
defining a $s$-analytic  function 
$C^{\infty}(\C\times {\cal M} \times {\cal M})$. This function gets 
the value $-P(x,y)$ for $s=0$. We can rearrange (\ref{dim}) after the 
analytic continuation of the term containing $P_{0}$
as
\begin{eqnarray}
\zeta(s,x,y| A/\mu^{2}) &=&  \zeta(N,s,x,y|A/\mu^{2}, \mu_{0}^{-2})
 - \left(\frac{\mu}{\mu_{0}} \right)^{2s}
 \frac{P_{0}(x,y)}{s\Gamma(s)}
 \nonumber\\
& &  +\frac{\mu^{2s}\chi (\sigma(x,y))}{(4\pi)^{D/2}\Gamma(s)}
 \sum_{j=0}^N a_j(x,y|A)\int_{0}^{\mu_{0}^{-2}} dt\: t^{s-1+j-D/2} 
{e^{-\sigma(x,y)/2t}}
  \label{dim'} 
 \end{eqnarray}
This can be considered  as another definition of 
$\zeta(N,s,x,y|A/\mu^{2}, \mu_{0}^{-2})$ equivalent to (\ref{s}) 
in the sense of the $s$ analytic continuation.
Concerning the last term of the right hand side of (\ref{dim'}) we 
have to distinguish between two cases. \\
For $x\neq y$, following procedures similar to those above, 
it is quite simply proven that
 the last term  in the right hand side 
defines an everywhere  $s$-analytic function 
$C^{\infty}(\C\times (({\cal M}\times{\cal M})-{\cal D}) )$ as it 
stands. Once again, 
this result is achieved essentially  because of the sharp decay 
of the exponential as $t\rightarrow 0^{+}$. We notice also that 
the considered term vanishes for $s=0$.\\
Summarizing, in the case $x\neq y$, the left hand side of (\ref{dim})
defines an everywhere $s$ analytic function which, at least,  belongs 
also to $C^{0}(\C\times (({\cal M}\times {\cal M}) -{\cal D}) )$
together
with all of its $s$ derivatives. Moreover it vanishes for $s=0$ giving 
rise
to (\ref{azero}) in the case $x\neq y$. The order of the zero at $s=0$
in the right hand side of (\ref{azero}) is at least $1$ because of the 
overall factor $1/\Gamma(s)$ in (\ref{dim'}). Up to now, we have proven 
$(a1),(a2)$ and $(c)$ partly.\\  
Let us finally consider the last term in the right hand side of (\ref{dim})
in the case $x=y$. In this case we cannot take advantage of the 
sharp   decay of the exponential. However, we can perform the 
integration for $Re$ $s > D/2 $ and then continue the result 
as far as it is possible in the remaining part of the $s$-complex plane.
Notice that, away from the poles, the obtained function is $C^{\infty}$
in $s,x,y$ trivially. We have finally 
\begin{eqnarray}
\zeta(s,x| A/\mu^{2}) &=& \zeta(N,s,x,x|A/\mu^{2}, \mu_{0}^{-2}) 
 -\frac{(\mu/\mu_{0})^{2s} P_{0}(x,x)}{s\Gamma(s)} \nonumber\\
 & & +  \frac{\mu^{2s}}{(4\pi)^{D/2}}
\sum_{j=0}^{N}
\frac{a_{j}(x,x|A) (\mu_{0}^{-2})^{(s+j-D/2)}}{\Gamma (s) (s+j-D/2)}
\label{dim2}\:.
\end{eqnarray}
This identity defines an analytic continuation 
 $\zeta(s,x| A/\mu^{2})$ at least 
 in ${\cal D}_{N} =\{  s\in \C | Re \mbox{ } s> D/2 - N \}$
for each integer $N > D/2+2$, indeed,
 therein  both functions (and all of their $s$ derivatives)
  in right hand side are defined and  continuous in $(x,s)$ away 
  from the possible poles. \\
Summarizing, the left  hand side of the equation above is decomposed into a 
function analytic in ${\cal D}_{N}$  and a function which is  
meromorphic in the same set, both functions and their $s$ derivatives 
are at least continuous in $(s,x)$ away from possible poles.
 Noticing that ${\cal D}_{N} \subset {\cal 
D}_{N+1}$ and $\bigcup_{N=1}^{+\infty} {\cal D}_{N} = \C$, 
the properties found out for the function $\zeta(s,x,y|A/\mu^{2})$ 
can be extended in the whole $s$-complex plane. 
In particular, the continued local $\zeta$ function and all of its $s$ derivatives  belong to 
  $C^{0}((\C-{\cal P})\times {\cal M})$ at least, where ${\cal P}$ 
  is the set of the actual 
poles of the last term in the right hand side of (\ref{dim2}).
Notice that, for $s=0$, (\ref{dim2}) gives (\ref{azero}) in the case 
$x=y$. This proves $(b)$ and complete the proof of $(c)$.\\
The proof of the part $(d)$ of the theorem concerning
the integrated $\zeta$ function is very similar to the 
case $x=y$ treated above. And one straightforwardly 
finds that the $s$-continuation procedure
commutes with the integration procedure of the  local (on-diagonal) $\zeta$
function as a consequence of the Fubini theorem.  $\Box$\\

\noindent {\bf Proof of Theorem 2.7.}
In our hypotheses, $Ker $ $B= \{ 0 \}$ and thus the local 
$\zeta$ function of $B$ is defined  as in (\ref{zetad}) for $x=y$,
 $Re$ $s>D/2$ and without the term $P(x,x|B)$ in the integrand.
The expression of $K(t,x,x|B)$ is very simple, in fact one has
\begin{eqnarray}
K(t,x,y|B) = e^{-\delta m^{2}t} K(t,x,y|A) \label{a2}\:.
\end{eqnarray}
Indeed, the right hand side satisfies trivially 
the heat equation (\ref{hk})  for $B$ whenever
$K(t,x,y|A)$ satisfies that equation for $A$, 
and this must be  the only solution because of  {\bf Theorem 1.1}.
Therefore, we have
\begin{eqnarray}
\zeta(s,x|B/\mu^{2}) = z_{1}(s,x) +z_{2}(s,x) + z_{3}(s,x) \label{tre}
\end{eqnarray}
where, we have  decomposed the right hand side in a 
sum of three parts after we have taken the expansion of the 
exponential $\exp {-\delta m^{2}t}$. For $Re$ $s>D/2$ they are,
 $P_{0}$ being the projector on the kernel 
of $A$ and  $\mu_{0}>0$ any real constant,
\begin{eqnarray}
z_{1}(s,x) &:= &\frac{\mu^{2s}}{\Gamma(s)}
\int_{\mu_{0}^{-2}}^{+\infty}dt\:\sum_{n=0}^{+\infty}\frac{(-\delta 
m^{2})^n}{n!} 
t^{s-1+n} [K(t,x,x|A) - P_{0}(x,x)] \label{az1}\\
z_{2}(s,x) &:= &\frac{\mu^{2s}}{\Gamma(s)}
\int_{0}^{\mu_{0}^{-2}} dt\:\sum_{n=0}^{+\infty}\frac{(-\delta m^{2})^n}{n!} 
t^{s-1+n} [K(t,x,x|A) - P_{0}(x,x)] \label{az2}\\
z_{3}(s,x) &:=& \left( \frac{\mu^{2}}{\delta m^{2}}\right)^s 
P_{0}(x,x)\:.
\label{az3}
\end{eqnarray}
The last term is the result of
\begin{eqnarray}
\frac{\mu^{2s}}{\Gamma(s)} \int_{0}^{+\infty} dt\:
e^{-\delta m^{2}t} 
t^{s-1} P_{0}(x,x) = \left( \frac{\mu^{2}}{\delta m^{2}}\right)^s 
P_{0}(x,x) \nonumber \:.
\end{eqnarray}
We want to study separately the $s$ analytic continuations of the
 first two terms above. The last term does not involves 
 particular problems.
  In particular we shall focus attention
 on the possibility of interchange the sum with the
 integral and we want to discuss the nature of the convergence of the series
 once one has interchanged the sum over $n$  with the  integral over 
 $t$.  
 In fact, we want to prove that the convergence is uniform in $s$ 
 within opportune sets.
 Let us start with $z_{1}(s,x)$. We consider the double integral
 in the measure "$n$-sum $\otimes$ $t$-integral" of 
 nonnegative elements
 \begin{eqnarray}
 I :=  \sum_{n=0}^{+\infty} \int_{\mu_{0}^{-2}}^{+\infty}dt \:
 \frac{(\delta m^{2})^n}{n!} 
|t^{s-1+n} [K(t,x,x|A) - P_{0}(x,x)]|
 \end{eqnarray}
 for $Re$ $s\in [\alpha, \beta]$, $\alpha,\beta \in \R$ fixed 
 arbitrarily, $t\in  [\mu_{0}^{-2}, +\infty)$.
 Following the same procedure than in the proof of {\bf Theorem 2.2},
 from (\ref{magg2})
 we have that, for any $\epsilon \in (0,1)$, there is a positive 
 $B$, such as, in the sets defined above 
 \begin{eqnarray}
\frac{(\delta m^{2})^n}{n!} 
\int_{\mu_{0}^{-2}}^{+\infty} dt\:
|t^{s-1+n}| |K(t,x,x|A) - P_{0}(x,x)| 
\leq a_{n} \:,\label{magga}
\end{eqnarray}
where 
\begin{eqnarray}
a_{n}:=
 \frac{(\delta m^{2})^n}{n!} 
\int_{\mu_{0}^{-2}}^{+\infty}dt\: t^n e^{-\epsilon \lambda t} 
\left( e^{B(\alpha-1)^{2}} + e^{B(\beta-1)^{2}}
\right)  \:.
 \end{eqnarray}
 The series of the positive  terms $a_{n}$ converges provided 
 $\epsilon \in (0,1)$ is chosen to give $\epsilon \lambda > \delta m^{2}$,
 the sum of $a_{n}$ being bounded by the $t$ integral, in 
 $[\mu^{2}_{0}, +\infty)$, of the function 
 \begin{eqnarray}
t \mapsto \left( e^{B(\alpha-1)^{2}} + e^{B(\beta-1)^{2}}\right)
 e^{-(\epsilon \lambda -\delta m^{2} )t}
 \end{eqnarray}
Therefore, by a part of Fubini's theorem,
 we have proven that  
 the function defined  for $Re$ $s$ fixed in $[\alpha, \beta]$, $t\in 
 [\mu^2_{0},+\infty)$, $n\in \N$
\begin{eqnarray}
(n,t) \mapsto \frac{(-\delta m^{2})^n}{n!} 
t^{s-1+n} [K(t,x,x|A) - P_{0}(x,x)]
 \end{eqnarray}
is integrable in the product measure above, and thus, again by Fubini's 
theorem, we can interchange the series with the integrals in (\ref{az1}).
Moreover the series of integrals so obtained is $s$-uniformly 
convergent in $Re$ $s\in [\alpha, \beta]$ because this series is 
term-by-term bounded by the series of the positive numbers 
$a_{n}$ which do not depend on $s$. The $s$-analyticity, for 
$s\in (\alpha, \beta)$,  
of the 
terms of the series of the integrals was still proven in the proof of {\bf 
Theorem 2.2}.  $z_{1}(s,x)$ defines an analytic function in any set 
$Re$ $s\in (\alpha, \beta)$ and it can be computed by summing the series 
of analytic functions, uniformly convergent
\begin{eqnarray}  
 z_{1}(s,x) = \sum_{n=0}^{+\infty}\frac{(-\delta 
m^{2})^n}{n!} \frac{\mu^{2s}}{\Gamma(s)}
\int_{\mu_{0}^{-2}}^{+\infty} dt\: 
t^{s-1+n} [K(t,x,x|A) - P_{0}(x,x)] \label{az1'}
\end{eqnarray}\\

Let us consider $z_{2}(s,x)$. By {\bf Theorem 1.4} for
$Re$ $s>D/2$, 
we can decompose 
\begin{eqnarray}
z_{2}(s,x) &=& \frac{\mu^{2s}}{\Gamma(s)}\int_{0}^{\mu^{-2}_{0}}
dt \sum_{n=0}^{+\infty}\frac{(-\delta m^{2})^n}{n!} t^{s-1+n-D/2+N}
O_\eta(t;x,x)\nonumber  \\
& + & \sum_{j=0}^{N}a'_{j}(x,x|A)\frac{\mu^{2s}}{\Gamma(s)}
\int_{0}^{\mu^{-2}_{0}}
dt \sum_{n=0}^{+\infty}\frac{(-\delta m^{2})^n}{n!} t^{s-1+n-D/2+j}\:.
\label{afine?}
\end{eqnarray}
 Above  $a'_{j}(x,x|A)=a'_{j}(x,x|A)/(4\pi)^{D/2}$ for $j\neq D/2$ and
 $a'_{D/2}(x,x|A)=a'_{D/2}(x,x|A)/(4\pi)^{D/2} - P_{0}(x,x)$ otherwise,
 moreover $N>D/2+2$ 
is any positive integer, and  $O_\eta(s;x,y)$ was defined in 
 the cited theorem (we have omitted the constant factor
 $(4\pi)^{-D/2}$). \\
 Let us  consider the first integral in the right hand side above.
 It is possible to   prove that, in any set $Re$ $s\in [D/2-N+\epsilon, 
 \beta]$, 
  $\beta\in \R$ ($N\geq D/2$ integer and $\epsilon >0$ are fixed 
 arbitrarily and $D/2-N+\epsilon < \beta$) one can 
  interchange the symbol of integration with that of series. Moreover
 the terms of the new series are analytic function of $s$ in 
 $Re$ $s\in (D/2-N+\epsilon, \beta)$
 and
 the new series converges $s$ uniformly in the considered closed set.\\
 The proof is very similar to the proof of the analogue 
 statement for $z_{1}(s,x)$. The the possibility to 
 interchange the integral with the series  follows from Fubini's theorem for the
 measure  "$n$-sum $\otimes$ $t$-integration"  exactly as in the 
 previous case,  and  employs the uniform bound, obtained from 
  (\ref{k'servi}) within 
 the proof of {\bf Theorem 2.2}
  \begin{eqnarray}
  \int_{0}^{\mu^{-2}_{0}} dt\: | t^{s-1+n-D/2} O_\eta(t;x;x) |
  \leq C \frac{\mu^{-2(n+\epsilon)}_{0}}{n+\epsilon}
  + C \frac{\mu^{-2(n+1)}_{0}}{n+1}
  \mu^{-2|\beta+N-D/2-1|}_{0} \label{abound}
   \end{eqnarray}
($C$ being  a positive constant). Notice also that, trivially,   
 \begin{eqnarray}
\sum_{n=0}^{+\infty}
  \frac{(\delta 
m^{2})^n}{n!}
   \left( \frac{\mu^{-2(n+\epsilon)}_{0}}{n+\epsilon}
  + \frac{\mu^{-2(n+1)}_{0}}{n+1}
  \mu^{-2|\beta+N-D/2-1|}_{0}\right) < +\infty
   \end{eqnarray}
 The analyticity of the terms in the  series of the integrals 
 deal with as pointed out  in the proof of {\bf Theorem 2.2}.
 The uniform convergence of the series follows from 
 the $s$-uniform bound (\ref{abound}).
 Hence, the  first term in the right hand side of (\ref{afine?})
 can be written as
 \begin{eqnarray}
z'_{2}(s,x) = \frac{\mu^{2s}}{\Gamma(s)}\sum_{n=0}^{+\infty}
\frac{(-\delta m^{2})^n}{n!} \int_{0}^{\mu^{-2}_{0}}
dt \: t^{s-1+n-D/2+N}
O_\eta(t;x,x) \label{afine?'}.
\end{eqnarray}
It defines an analytic function in any set $Re$ $s \in [D/2-N+\epsilon, 
\beta]$
and the convergence of the series is uniform therein. Notice that
$N$ as well as $\beta$ are arbitrary and thus the set above can be 
enlarged arbitrarily.

 Finally let us consider the remaining term in the right hand side of 
 (\ref{afine?}) initially defined for $Re$ $s>D/2$ 
 \begin{eqnarray}
z''_{2}(s,x) &=& 
\sum_{j=0}^{N}a'_{j}(x,x|A)\frac{\mu^{2s}}{\Gamma(s)}\int_{0}^{\mu^{-2}_{0}}
dt \sum_{n=0}^{+\infty}\frac{(-\delta m^{2})^n}{n!} t^{s-1+n-D/2+j}\:.
\label{afine!}
\end{eqnarray}
 We can decompose the sum over $n$ into two parts 
 \begin{eqnarray}
z''_{2}(s,x) &=&  \sum_{n=0}^{K}\frac{(-\delta m^{2})^n}{n!}
\sum_{j=0}^{N}a'_{j}(x,x|A)\frac{\mu^{2s}}{\Gamma(s)}\int_{0}^{\mu^{-2}_{0}}
dt\: t^{s-1+n-D/2+j}
\nonumber\\
&+& 
\sum_{j=0}^{N}a'_{j}(x,x|A)\frac{\mu^{2s}}{\Gamma(s)}\int_{0}^{\mu^{-2}_{0}}
dt\:\sum_{n=K+1}^{+\infty}\frac{(-\delta m^{2})^n}{n!} 
t^{s-1+n-D/2+j}\:.
\label{afine!'}
\end{eqnarray}
 The first part can be continued in the whole $s$ complex plane
 away from the poles obtained by executing the integrals. The 
second part can be studied as in the cases discussed  above.
One has to study the convergence of 
\begin{eqnarray}
I' := \sum_{n=K+1}^{+\infty}\frac{(\delta m^{2})^n}{n!} 
\sum_{j=0}^{N}|a'_{j}(x,x|A)|\int_{0}^{\mu^{-2}_{0}}
dt\: |t^{s-1+n-D/2+j}|\:.
\end{eqnarray}
Let us consider $Re$ $s\in [D/2-K, \beta]$, where $\beta\in \R$ is 
arbitrary. Let us pose 
\begin{eqnarray}
 M := \sup_{s\in [D/2-K, \beta], 0\leq j \leq N} 
  \left\{  |a'_{j}(x,x|A)| \mu^{-2(Re s -D/2 +j)}_{0} \right\}\:,
\end{eqnarray}
 then we have (notice that $n= K+1, K+2,\ldots$)
 \begin{eqnarray}
 \frac{(\delta m^{2})^n}{n!} 
\sum_{j}^{N}|a'_{j}(x,x|A)|\int_{0}^{\mu^{-2}_{0}}
dt\: |t^{s-1+n-D/2+j}|  \leq \frac{(\delta m^{2})^n}{n!} 
   \frac{(N+1) M \mu^{-2n}_{0}}{n - K} =: b_{n}\:.
 \end{eqnarray}
The series of the positive coefficients  $b_{n}$ converges trivially.
As in the  cases previously treated, this is sufficient to assure
the possibility to interchange the sum over $n$ with the $t$ 
integration in the second line of (\ref{afine!'}) and assure the 
$s$-uniform convergence of the consequent series in $Re$ $s\in [D/2-K, \beta]$.
This completes  the proof of the  theorem. $\Box$.

 \end{document}